\documentclass{article}

\usepackage[T1]{fontenc}
\usepackage{soul} 
\usepackage{phfqit}
\usepackage{proba} 
\usepackage{amssymb}
\usepackage{amsmath}
\usepackage{amsthm}
\usepackage{amsfonts}
\usepackage{mathtools}
\usepackage{xspace}
\usepackage{bm}
\usepackage{nicefrac}
\usepackage{commath}
\usepackage{bbm}
\usepackage{boxedminipage}
\usepackage{xparse}
\usepackage{xcolor}
\usepackage{float}
\usepackage{multirow}
\usepackage{graphicx}
\usepackage{caption}
\usepackage{subcaption}
\usepackage{ifthen}
\usepackage{algpseudocode}
\usepackage{colortbl} 
\usepackage{fancyhdr}   
\usepackage{hyperref}
    \hypersetup{ colorlinks=true, linkcolor=blue, filecolor=magenta, urlcolor=red,
    citecolor=blue,}
\usepackage{fullpage}
\usepackage[utf8]{inputenc}
\usepackage{environ}
\usepackage[colorinlistoftodos]{todonotes}
\usepackage{tabu}
\usepackage{mdframed}

\newcommand{\jeremiah}[1]{\textcolor{red}{{\bf (Jeremiah:} {#1}{\bf ) }} \marginpar{\tiny\bf
             \begin{minipage}[t]{0.5in}
               \raggedright J:
            \end{minipage}}}
            
\newtheorem{proposition}{Proposition}
\newtheorem{lemma}{Lemma}
\newtheorem{theorem}{Theorem}
\newtheorem{corollary}{Corollary}
\newtheorem{definition}{Definition}

\newtheorem{claim}{Claim}

\usepackage{amsthm,thmtools}%
\theoremstyle{definition}

\theoremstyle{plain}

\newcommand{\etal}{\text{et al.}\xspace}

\newcommand{\polylog}{\ensuremath{\mathrm{polylog}}\xspace}
\newcommand{\negl}{\ensuremath{\mathsf{negl}}\xspace}
\newcommand{\secpar}{\ensuremath{{\kappa}}\xspace}
\newcommand{\zo}{\ensuremath{{\{0,1\}}}\xspace}

\newcommand{\eps}{\ensuremath{\varepsilon}}

\NewEnviron{problem}[1]{%
	\begin{center}\fbox{\parbox{6in}{%
				{\centering\scshape #1\par}%
				\parskip=1ex
				\everypar{\hangindent=1em}%
				\BODY
}}\end{center}}

\newcommand{\safe}{\ensuremath{\mathbb{S}_\mathbb{C}}}

\newcommand{\nbad}{\overline{\mathsf{bad}}}

\newcommand{\resspace}{\ensuremath{\mathsf{space}}}
\newcommand{\resspacet}{\ensuremath{\mathsf{space-time}}}
\newcommand{\restime}{\ensuremath{\mathsf{time}}}
\newcommand{\resst}{\ensuremath{\mathsf{ST}}}
\newcommand{\rescmc}{\ensuremath{\mathsf{CMC}}}
\newcommand{\rescq}{\ensuremath{\mathsf{CQ}}}
\newcommand{\res}{\ensuremath{\mathcal{R}}}
\newcommand{\constraints}{\ensuremath{\mathcal{M}}\xspace}
\newcommand{\class}{\ensuremath{\mathbb{C}}\xspace}

\newcommand{\channel}{\ensuremath{\mathcal{C}}\xspace}
\newcommand{\sender}{\ensuremath{\mathcal{S}}\xspace}
\newcommand{\receiver}{\ensuremath{\mathcal{R}}\xspace}
\newcommand{\msg}{\ensuremath{x}\xspace}
\newcommand{\codeword}{\ensuremath{y}\xspace}
\newcommand{\corrcodeword}{\ensuremath{y'}\xspace}
\newcommand{\lenalphabet}{\ensuremath{q}\xspace}
\newcommand{\lenmsg}{\ensuremath{k}\xspace}
\newcommand{\lencodeword}{\ensuremath{K}\xspace}
\newcommand{\alphabet}{\ensuremath{\Sigma}\xspace}
\newcommand{\encoder}{\ensuremath{\mathsf{Enc}}}
\newcommand{\decoder}{\ensuremath{\mathsf{Dec}}}
\newcommand{\loc}{\ensuremath{\ell}\xspace}
\newcommand{\errorrate}{\ensuremath{\rho}\xspace}
\newcommand{\probrecov}{\ensuremath{p}\xspace}

\newcommand{\superscript}[1]{\ensuremath{^{\mbox{\tiny{\textit{#1}}}}}\xspace}
\def \th {\superscript{th}}     
\def \etal{\,{\it et~al.}\,}
\newcommand{\ignore}[1]{{}}

\newcommand{\PPr}[1]{\ensuremath{\mathbf{Pr}\left[#1\right]}}

\renewcommand{\O}[1]{\ensuremath{\mathcal{O}\left(#1\right)}}

\newcommand{\pROM}{\textsf{pROM}\xspace}
\newcommand{\etrace}{\ensuremath{\mathsf{Trace}_{\mathcal{A},R,\mathsf{H}}}}

\newcommand{\PPT}{\textsf{PPT}}

\newcommand{\bad}{\textsf{bad}}

\newcommand{\indeg}{\ensuremath{\mathsf{indeg}}}

\newcommand{\sinks}{\ensuremath{\mathsf{sinks}}}
\newcommand{\parents}{\ensuremath{\mathsf{parents}}}
\newcommand{\HIGH}{\ensuremath{\mathsf{HIGH}}}

\newcommand{\dlab}{\ensuremath{\mathsf{lab}}}
\newcommand{\lab}{\dlab}
\newcommand{\majority}{\ensuremath{\mathsf{majority}}\xspace}
\newcommand{\spacepeb}{\ensuremath{\mathsf{space}}}
\newcommand{\mcost}{\ensuremath{\mathsf{mcost}}}

\newcommand{\HAM}{\ensuremath{\mathsf{HAM}}}
\newcommand{\attack}{\ensuremath{\mathcal{A}^{H(\cdot)}}}
\newcommand{\pPeb}{\Peb^{\parallel}}
\newcommand{\Peb}{{\cal P}} 

\newcommand{\oracleH}{\ensuremath{\mathsf{H}(\cdot)}}
\newcommand{\randOracleH}[1]{\ensuremath{\mathsf{H}(#1)}}

\newcommand{\attacker}{\ensuremath{\mathcal{A}}}
\newcommand{\roattacker}{\ensuremath{\mathcal{A^{\oracleH}}}}

\newcommand{\Enc}{\ensuremath{\mathsf{Enc}}}
\newcommand{\Dec}{\ensuremath{\mathsf{Dec}}}
\newcommand{\genkey}{\ensuremath{\mathsf{GenKey}}\xspace}
\newcommand{\sk}{\ensuremath{\mathsf{sk}}\xspace}
    \newcommand{\justesendec}{\ensuremath{\mathsf{Dec}_\mathsf{J}}\xspace}
    \newcommand{\justesenenc}{\ensuremath{\mathsf{Enc}_\mathsf{J}}\xspace}

    \newcommand{\Encjrep}{\ensuremath{\mathsf{Enc}_\mathsf{JREP}}\xspace}
    \newcommand{\Decjrep}{\ensuremath{\mathsf{Dec}_\mathsf{JREP}}\xspace}
    \newcommand{\rhojrep}{\ensuremath{\rho_\mathsf{JREP}}\xspace}
    
    \newcommand{\opskey}{\ensuremath{\mathsf{GenKey}_{\mathsf{OPS}}}\xspace}
    \newcommand{\opsenc}{\ensuremath{\mathsf{Enc}_{\mathsf{OPS}}}\xspace}
    \newcommand{\opsdec}{\ensuremath{\mathsf{Dec}_{\mathsf{OPS}}}\xspace}
    \newcommand{\opsrho}{\ensuremath{\mathcal{\rho}_{\mathsf{OPS}}}\xspace}
    \newcommand{\opsquery}{\ensuremath{\mathsf{\ell}_{\mathsf{OPS}}}\xspace}
    
    \newcommand{\Encro}{\ensuremath{\mathsf{Enc}^{\oracleH}_{\mathsf{final}}}}
    \newcommand{\Decro}{\ensuremath{\mathsf{Dec}^{\oracleH}_{\mathsf{final}}}}

\newcommand{\hybridusenc}{\ensuremath{\mathsf{Enc}_{0}^{\oracleH}\xspace}}
\newcommand{\distinguisher}{\ensuremath{\mathcal{D}}\xspace}
\newcommand{\ADpair}{\ensuremath{(\roattacker, \distinguisher)}\xspace}
\newcommand{\advantage}{\ensuremath{\mathsf{Adv}}\xspace}
\newcommand{\advantageAD}{\ensuremath{\advantage_{\attacker,\distinguisher}}\xspace}

\newcommand{\FAIL}{\ensuremath{\mathsf{fail}}\xspace}
\newcommand{\KEY}{\ensuremath{\mathsf{key}}\xspace}
\newcommand{\nKEY}{\ensuremath{\overline{\mathsf{key}}}\xspace}

\newcommand{\controlinfo}{\ensuremath{\mathsf{r}}\xspace}

\newcommand{\jcontrolcode}{\ensuremath{\mathsf{c}_{\mathsf{J}}\xspace}}
\newcommand{\lenjcontrolcode}{\ensuremath{\mathsf{L}_{\mathsf{J}}\xspace}}

\newcommand{\jrepcontrolcode}{\ensuremath{\mathsf{C}_\mathsf{JREP}\xspace}}
\newcommand{\numjrepcontrolcode}{\ensuremath{\mathsf{n}_{\mathsf{JREP}}}\xspace}
\newcommand{\opscode}{\mathsf{C}_\mathsf{OPS}\xspace}

\newcommand{\finalsk}{\ensuremath{\sk_{\mathsf{FINAL}}}\xspace}

\newcommand{\jrepquery}{\ensuremath{\ell_{\mathsf{JREP}}}\xspace}
\newcommand{\lenjrep}{\ensuremath{\mathsf{L}_\mathsf{JREP}}\xspace}

\newcommand{\rhoj}{\ensuremath{\rho_{\mathsf{J}}}\xspace}

\newcommand{\LDC}{\ensuremath{\mathsf{LDC}}\xspace}
\newcommand{\LDCs}{\ensuremath{\mathsf{LDC}}s}

\newcommand{\RLDCs}{\ensuremath{\mathsf{RLDC}}s}
\newcommand{\RLCCs}{\ensuremath{\mathsf{RLCC}}s}
\newcommand{\SUCCESS}{\ensuremath{\mathsf{succ}}\xspace}
\newcommand{\nSUCCESS}{\ensuremath{\overline{\mathsf{succ}}}\xspace}
\newcommand{\myindent}{\hspace*{6mm} }

\providecommand{\email}[1]{\href{mailto:#1}{\nolinkurl{#1}\xspace}}

\newcommand{\parameters}[1]{\ensuremath{{\color{gray}[#1]}}\xspace}

\newcommand{\codePriv}{\ensuremath{\mathsf{C_{\mathsf{priv}}}}\xspace}
\newcommand{\encPriv}{\ensuremath{\Enc_{\mathsf{priv}}}\xspace}
\newcommand{\decPriv}{\ensuremath{\Dec_{\mathsf{priv}}}\xspace}
\newcommand{\genkeyPriv}{\ensuremath{\mathsf{GenKey}_{\mathsf{priv}}}\xspace}
\newcommand{\lenmsgPriv}{\ensuremath{\lenmsg_{\mathsf{priv}}}\xspace}
\newcommand{\lencodewordPriv}{\ensuremath{\lencodeword_{\mathsf{priv}}}\xspace}
\newcommand{\locPriv}{\ensuremath{\loc_{\mathsf{priv}}}\xspace}
\newcommand{\errorratePriv}{\ensuremath{\errorrate_{\mathsf{priv}}}\xspace}
\newcommand{\probrecovPriv}{\ensuremath{\probrecov_{\mathsf{priv}}}\xspace}
\newcommand{\epsilonPriv}{\ensuremath{\epsilon_{\mathsf{priv}}}\xspace}
\newcommand{\codewordPriv}{\ensuremath{Y_{\mathsf{priv}}}\xspace}

\newcommand{\codeLDC}{\ensuremath{\mathsf{C_{\mathsf{ldc^*}}}}\xspace}
\newcommand{\encLDC}{\ensuremath{\Enc_{\mathsf{ldc^*}}}\xspace}
\newcommand{\decLDC}{\ensuremath{\Dec_{\mathsf{ldc^*}}}\xspace}
\newcommand{\lenmsgLDC}{\ensuremath{\lenmsg_{\mathsf{ldc^*}}}\xspace}
\newcommand{\lencodewordLDC}{\ensuremath{\lencodeword_{\mathsf{ldc^*}}}\xspace}
\newcommand{\locLDC}{\ensuremath{\loc_{\mathsf{ldc^*}}}\xspace}
\newcommand{\errorrateLDC}{\ensuremath{\errorrate_{\mathsf{ldc^*}}}\xspace}
\newcommand{\probrecovLDC}{\ensuremath{\probrecov_{\mathsf{ldc^*}}}\xspace}
\newcommand{\codewordLDC}{\ensuremath{Y_{\mathsf{ldc^*}}}\xspace}

\newcommand{\codeFinal}{\ensuremath{\mathsf{C_{\mathsf{final}}}}\xspace}
\newcommand{\encFinal}{\ensuremath{\Enc^{\oracleH}_{\mathsf{final}}}\xspace}
\newcommand{\decFinal}{\ensuremath{\Dec^{\oracleH}_{\mathsf{final}}}\xspace}
\newcommand{\lenmsgFinal}{\ensuremath{\lenmsg_{\mathsf{final}}}\xspace}
\newcommand{\lencodewordFinal}{\ensuremath{\lencodeword_{\mathsf{final}}}\xspace}
\newcommand{\locFinal}{\ensuremath{\loc_{\mathsf{final}}}\xspace}
\newcommand{\errorrateFinal}{\ensuremath{\errorrate_{\mathsf{final}}}\xspace}
\newcommand{\probrecovFinal}{\ensuremath{\probrecov_{\mathsf{final}}}\xspace}
\newcommand{\epsilonFinal}{\ensuremath{\epsilon_{\mathsf{final}}}\xspace}
\newcommand{\skFinal}{\ensuremath{\sk_{\mathsf{final}}}\xspace}

\newcommand{\code}{\ensuremath{{\mathsf{C}}}\xspace}

\newcommand{\encHybridZero}{\ensuremath{\Enc^{\oracleH}_{0}}\xspace}
\newcommand{\seedHybridZero}{\ensuremath{\mathsf{r}^{(0)}}\xspace}
\newcommand{\randomnessHybridZero}{\ensuremath{\mathsf{R}^{(0)}}\xspace}
\newcommand{\skHybridZero}{\ensuremath{\mathsf{sk}^{(0)}}\xspace}
\newcommand{\ldccodewordHybridZero}{\ensuremath{\mathsf{Y_{ldc}}^{(0)}}\xspace}
\newcommand{\privcodewordHybridZero}{\ensuremath{\mathsf{Y_{priv}}^{(0)}}\xspace}
\newcommand{\codewordHybridZero}{\ensuremath{\mathsf{Y_{hyb}}^{(0)}}\xspace}
\newcommand{\codeHybridZero}{\ensuremath{\mathsf{C}_0}\xspace}
\newcommand{\epsilonHybridZero}{\ensuremath{\epsilon_0}\xspace}

\newcommand{\encHybridOne}{\ensuremath{\Enc^{\oracleH}_{1}}\xspace}
\newcommand{\decHybridOne}{\ensuremath{\Dec^{\oracleH}_{\mathsf{priv}^{*}}}\xspace}
\newcommand{\seedHybridOne}{\ensuremath{\mathsf{r}^{(1)}}\xspace}

\newcommand{\skHybridOne}{\ensuremath{\mathsf{sk}^{(1)}}\xspace}
\newcommand{\ldccodewordHybridOne}{\ensuremath{\mathsf{Y_{ldc^*}}^{(1)}}\xspace}
\newcommand{\privcodewordHybridOne}{\ensuremath{\mathsf{Y_{priv}}^{(1)}}\xspace}
\newcommand{\codewordHybridOne}{\ensuremath{\mathsf{Y_{hyb}}^{(1)}}\xspace}
\newcommand{\codeHybridOne}{\ensuremath{\mathsf{C}_1}\xspace}
\newcommand{\epsilonHybridOne}{\ensuremath{\epsilon_1}\xspace}

\newcommand{\indistinguishExp}{\ensuremath{\mathtt{Exp}_{\attacker,\distinguisher,\mathsf{H},\kappa,x}}\xspace}

\newcommand{\encHybridExp}{\ensuremath{\Enc^{\oracleH}_{b}}\xspace}
\newcommand{\seedHybridExp}{\ensuremath{\mathsf{r}^{(b)}}\xspace}

\newcommand{\skHybridExp}{\ensuremath{\mathsf{sk}^{(b)}}\xspace}
\newcommand{\ldccodewordHybridExp}{\ensuremath{\mathsf{Y_{ldc^*}}^{(b)}}\xspace}
\newcommand{\privcodewordHybridExp}{\ensuremath{\mathsf{Y_{priv}}^{(b)}}\xspace}
\newcommand{\codewordHybridExp}{\ensuremath{\mathsf{Y_{hyb}}^{(b)}}\xspace}
\newcommand{\corrcodewordHybridExp}{\ensuremath{\mathsf{Y_{hyb}}^{(b)'}}\xspace}

\newcommand{\numOracleQueries}{\ensuremath{q}}

\newcommand{\LDCvariant}{\ensuremath{\LDC^*}}
\newcommand{\LDCsvariant}{\ensuremath{\LDC^*\text{s}}}

\newcommand{\LDCsecgame}{\ensuremath{\mathtt{{LDC-Sec-Game}}\parameters{\attacker,\msg, \mathsf{H}, \secpar, \errorrate, \probrecov}}\xspace}

\newcommand{\privLDCsecgameHybrid}{\ensuremath{\mathtt{{priv-LDC-Sec-Game}}\parameters{\attacker,\msg, \secpar, \errorrateFinal, \probrecovFinal}}\xspace}

\title{
On Locally Decodable Codes in Resource Bounded Channels}

\author{Jeremiah Blocki\thanks{Department of Computer Science, Purdue University, West Lafayette, IN. 
Email: \email{jblocki@purdue.edu}. }
\and
Shubhang Kulkarni\thanks{Department of Computer Science, Purdue University, West Lafayette, IN.
Email: \email{kulkar17@purdue.edu}}
\and
Samson Zhou\thanks{School of Computer Science, Carnegie Mellon University, Pittsburgh, PA. 
E-mail: \email{samsonzhou@gmail.com}}}
\date{\today}

\begin{document}
\maketitle

\begin{abstract}
Constructions of locally decodable codes (\LDCs) have one of two undesirable properties: low rate or high locality (polynomial in the length of the message). 
In settings where the encoder/decoder have already exchanged cryptographic keys 
and the channel is a probabilistic polynomial time (\PPT) algorithm, it is possible to circumvent these barriers and design \LDCs\ with constant rate and small locality. 
However, the assumption that the encoder/decoder have exchanged cryptographic keys is often prohibitive. 
We thus consider the problem of designing explicit and efficient \LDCs\ in settings where the channel is {\em slightly} more constrained than the encoder/decoder with respect to some resource e.g., space or (sequential) time. 
Given an explicit function $f$ that the channel cannot compute, we show how the encoder can transmit a random secret key to the local decoder using $f(\cdot)$ and a random oracle $\oracleH$. 
We then bootstrap the private key \LDC\ construction of Ostrovsky, Pandey and Sahai (ICALP, 2007), thereby answering an open question posed by Guruswami and Smith (FOCS 2010) of whether such bootstrapping techniques are applicable to \LDCs\ in channel models weaker than just \PPT\ algorithms. 
Specifically, in the random oracle model we show how to construct explicit constant rate \LDCs\ with locality of $\polylog$ in the security parameter against various resource constrained channels.

\end{abstract}
\section{Introduction}

\myindent
Consider the classical one-way communication setting where two parties, the \emph{sender} and \emph{receiver}, communicate over a \emph{noisy channel} that may \emph{corrupt} parts of any message sent over it. An \emph{error correcting code} is an invertible transformation mapping messages into \emph{codewords} that are then transmitted over the noisy channel. The goal is to ensure that the decoder can (w.h.p.) reliably recover the entire message from the corrupted codeword. For locally decodable codes (\LDCs) we have an even stronger goal: The decoder should be able to reliably recover \emph{any} individual bit of the original message (w.h.p.) by examining at most $\ell$ bits of the corrupted codeword. An ideal \LDC{} should have a good rate (i.e., the codeword should not be much longer than the original message) and small locality $\ell$. 

\myindent Historically, there have been two major lines of work associated with modelling the channel behavior. In Shannon's \emph{symmetric channel} model, the channel corrupts each bit of the codeword independently at random with some fixed probability. By contrast, in Hamming's \emph{adversarial channel} model the channel corrupts the codeword in a worst case manner subject to an upper bound on the total number of corruptions.

\myindent Unsurprisingly, when we work in Shannon's channel model it is much easier to design  \LDCs\ with good rate/locality. By contrast, state of the art \LDC{} constructions for Hamming channels either have very high locality e.g., $\ell = 2^{\O{\sqrt{\log n \log \log n}}}$ \cite{KoppartyMRS17} or poor rate e.g., Hadamard codes have constant locality $\ell = \O{1}$ but the codeword has exponential length. Unfortunately, in many real-world settings independent random noise is not a realistic model of channel behavior e.g., burst-errors are common in reality, but unlikely in Shannon's model. Thus, coding schemes designed to work in Shannon's channel model are not necessarily suitable in practice. By contrast, coding schemes designed to work in Hamming's adversarial setting must be able to handle \emph{any} error pattern. 

\myindent Our central motivating goal is to find classes of adversarial channels that are expressive enough to model any error patterns that would arise in nature, yet admit \LDCs{} with good decoding algorithms. \LDCs\ have found remarkable applications throughout various fields, notably private information retrieval schemes \cite{private_info_retrieval_1, private_info_retrieval_2_Chor, private_info_retrieval_3}, psuedo-random generator constructions \cite{psuedo_random_gen_1, psuedo_random_gen_2}, self-correcting computations \cite{self_correction1, self_correction_2_Gemmel}, \textsf{PCP} systems \cite{Babai_ldc} and fault tolerant storage systems \cite{fault_tolerant_storage}.  

\myindent Lipton~\cite{lipton_new_1994} introduced the \emph{adversarial computationally bounded} model, where the channel was viewed as a Hamming channel restricted to bounded corruption by a \emph{probabilistic polynomial time} (\PPT) algorithm. The notion of adversaries being computationally bounded is well-motivated by real-world channels that have some sort of limitations on their computations i.e., we expect error patterns encountered in nature to be modeled by some (possibly unknown) \PPT{} algorithm. We argue that even Lipton's channel significantly overestimates the capability of the channel. For example, if the channel has reasonably small latency, say $10$ seconds, and the world's fastest single core processor can evaluate 10 billion instructions per second then the depth of any (parallel) computation performed by the channel is at most $100$ billion operations. 

\myindent 
This view of modelling the channel as more restricted than just \PPT\ was further explored by Guruswami and Smith~\cite{Guruswami_Smith:2016} who studied channels that could be described by simple (low-depth) circuits. 
Remarkably, even such a simple restriction allowed them to design codes that enjoyed no public/private key setup assumptions, while matching the Shannon capacity using polynomial time encoding/decoding algorithms. With such positive results, it is natural to ask whether similar results may be expected for \LDCs. 

\subsection{Contributions}
\myindent 
We introduce \emph{resource bounded adversarial channel} models which admit \LDCs\ with good locality whilst still being expressive enough to plausibly capture any error pattern for most real-world channels. We argue that these resource bounded channel models are already sufficiently expressive to model any corruption pattern that might occur in nature e.g., burst-errors, correlated errors. For example, observe that the channel must compute the entire error pattern \emph{before} the codeword is delivered to the receiver. Thus, the channel can be viewed as \emph{sequentially time bounded} e.g., the channel may  perform arbitrary computation in parallel but the total depth of computation is bounded by the latency of the channel. The notion of a space bound (or space-time bound) channel can be similarly motivated.

\myindent We introduce \emph{safe functions} as a general way to characterize \LDC friendly channels. Intuitively, a function $f$ is ``safe'' for a class of channels if the channel is not able to predict $f(x)$ given $x$. We show how to construct safe functions for several classes of resource bounded channels including time bounded, space bounded, and cumulative memory cost bounded channels in the parallel random oracle model. For example, in the random oracle model the function $\mathsf{H}^{t+1}(x)$ is a safe function for the class of sequentially time-bounded adversaries i.e., it is not possible to evaluate the function using fewer than $t$ sequential calls to the random oracle $\mathsf{H}$. We also discuss how to construct safe functions for the class of space (resp. space-time) bounded channels using random oracles. 




\myindent Furthermore, we give a general framework for designing good locally decodable codes against resource bounded adversarial channels by using safe functions to bootstrap existing private-key \LDC\ constructions. Our framework assumes no a priori private or public key setup assumptions, and constructs explicit \LDCs\ over the binary alphabet\footnote{Note that small alphabet sizes are attractive for practical channels designed to transmit bits efficiently.} with constant rate against \emph{any} class of resource bounded adversaries admitting \emph{safe functions}.

\myindent 
Our local decoder can decode correctly with arbitrarily high constant probability after examining at most $\O{f(\secpar)}$ bits of the corrupted codeword, where \secpar is the security parameter\footnote{In this paper we use the security parameter $\secpar$ in an asymptotic sense e.g., for any attacker running in time $\mathtt{poly}(\secpar)$ there is a negligible function $\mathtt{negl}(\secpar)$ upper bounding the probability that the attacker succeeds. 
In particular, the function $\mathtt{negl}(\secpar)=2^{-\log^{1+\eps} \secpar)}$ is negligible, but does not provide $\secpar$-bits of concrete security i.e., any attacker running in time $t$ succeeds with probability at most $t2^{-\secpar}$.} and $f(\secpar)$ is any function such that $f(\secpar) = \omega(\log\secpar)$ e.g., $f(\secpar) = \log^{1+\eps} \kappa$ or $f(\secpar) = \log \secpar \log \log \secpar$. 
By contrast, state of the art \LDC{} constructions for Hamming channels have very high locality e.g., $2^{\O{\sqrt{\log n \log \log n}}}$ \cite{KoppartyMRS17}. 
Our codes are robust against a constant fraction of corruptions, and are (essentially) \emph{non-adaptive} i.e., the local decoding algorithm can decode after submitting just two batches of queries. 

\myindent 
Our constructions stand at the intersection of coding theory and cryptography, using well-known tools and techniques from cryptography to provide notions of (information theoretic) randomness and security for communication protocols between sender/receiver. 
To prove the security of our constructions, we introduce a \emph{two-phase distinguisher hybrid argument}, which may be of independent interest for other coding theoretic problems in these resource bounded channel models.

\subsection{Technical Overview}


\paragraph{Private \LDCs.} 
Our starting point is the private locally decodable codes of \cite{OPS}. 
These \LDCs{} permit nearly optimal query complexity, asymptotically positive rate and reliable decoding with high probability, but make the strong assumption that the sender and receiver have already exchanged a secret key $K$ that is unknown to the \PPT\ adversarial channel over which they communicate. In our setting the sender and the receiver do not have access to any secret key. Our constructions thus \emph{reduce} the general setting (no setup assumptions) against resource bounded channels to the shared private key setting against these channels, so that we can bootstrap private \LDC{} constructions.

\paragraph{Bootstrapped Encoder/Decoder.}
Our encoder uses the following high level  template: (1) samples a random seed $r$ (2) computes a predetermined safe function $f(r)$ on the seed and extracts a secret key $K$ from $f(r)$ (e.g., using a random oracle) (3) Uses the private \LDC \ encoder to encode the message using $K$ (4) appends a reliable encoding (repetition code) of the random seed $r$ to the codeword. The local decoder (1) decodes the random seed $r$ (random sampling + majority vote). (2) Evaluates the safe function $f(r)$ to recover the secret key $K$. (3) Uses the private \LDC \ decoder with the secret key $K$ to recover the desired bit of the original message. 

\paragraph{Security Proof.} We remark that there are a few subtle challenges that arise when we prove that our bootstrapped construction is secure. We want to prove that the channel will (w.h.p.) fail to produce a corrupted codeword that fools the local decoding algorithm. Towards this goal we might try to prove that the channel cannot distinguish the derived key $K$ from a truly random key even given the nonce $r$. However, this is insufficient to prove that the local decoder is successful because the local decoder \emph{is} able to recover $K$ from $f$. We introduce a novel \emph{two-phase distinguisher game} to address these challenges. In particular, we consider an attacker-distinguisher pair who tries to predict whether or not the secret encoding key $K$ is derived from the nonce $r$ $(b=0)$ or was selected uniformly at random $(b=1)$. In phase 1 the (resource bounded) attacker generates a corrupted codeword which is given to the distinguisher in phase 2 who must then guess   whether $b =1$ or $b=0$. The  distinguisher is computationally unbounded, but is not allowed to query the random oracle. If $f$ is a safe function then the advantage of any such attacker-distinguisher pair can be shown to be negligible. We  demonstrate that any channel which succeeds at fooling our local decoder yields an attacker-distinguisher pair for this two phase game --- the distinguisher works by simulating the private \LDC\ decoder to distinguish between the two aforementioned encodings. It follows that the channel cannot fool the local decoder (except with negligible probability).

\subsection{Related Work} \label{sec:relatedwork}
\myindent 
Many existing code constructions consider an underlying channel that can only introduce a bounded number of errors, but has an unlimited time to adversarially decide the positions of these errors.  
These codes are therefore resilient to any possible error pattern with a bounded number of corruptions, corresponding to Hamming's error model, and are safe for data transmission. 
However, this resiliency to the worst-case error leads to coding limitations and some possibly undesirable tradeoffs. 
On one hand, current constructions for \LDCs\ that focus on efficient encoding can obtain any constant rate $R<1$ while simultaneously being robust to any constant fraction $\delta<1-R$ of errors and using $2^{\O{\sqrt{\log n \log \log n}}}$ queries for decoding \cite{KoppartyMRS17}. 
On the other hand, codes that focus on low query complexity obtain blocklength that is subexponential in the message length while using a constant number of queries $q\le 3$~\cite{Yekhanin08,Efremenko12,DvirGY11}. 
Finally, if exactly $q=2$ queries are desired, any code \emph{must} use blocklength exponential in the message length~\cite{KerenidisW04}. 
Avoiding such drastic tradeoffs between blocklength and query complexity would be attractive for other natural channels in contrast to Hamming's error model. 
For example, Shannon introduces a model in which each symbol has some independent probability of being corrupted; this probability is generally fixed across all symbols and known a priori.  
However, this probabilistic channel may be too weak to capture natural phenomenon such as bursts of consecutive error. 

\myindent 
Thus it is reasonable to believe that many natural channels lie between these two extremes; in particular, Lipton~\cite{lipton_new_1994} argues that many reasonable channels are computationally bounded and can be modeled as \PPT{} algorithms. 
In this model, \cite{lipton_new_1994} introduced an analog to classical error-correcting codes that is robust to a fraction of errors beyond the rates provably tolerable by \emph{any} code in the adversarial Hamming channel model. 
Similarly, a line of work~\cite{lipton_new_1994,micali_etal_computationally_bounded_noise,Guruswami_Smith:2016, ShaltielS16} have improved upon the error rate limits of classical error-correcting codes in slight variants of Lipton's computationally bounded channel model. 
A weakness of the codes introduced by~\cite{lipton_new_1994} is the strong cryptographic assumption that the sender and receiver share a {\em secret} random string unknown to the channel. 
This weakness is ameliorated by \cite{micali_etal_computationally_bounded_noise}, who observe that if a message is encoded by digitally signing a code that is \emph{list-decodable} with a secret key, then an adversarial \PPT{} is unlikely to produce valid signatures. 
Conversely, the decoder can select the unique message from the list of possible messages with a valid signature, effectively producing public-key error-correcting codes against computationally bounded channels. 
Subsequently, \cite{Guruswami_Smith:2016} further removes the public-key setup assumption specifically for the channel in which either the error is independent of the actual message being sent, or the errors can be described by polynomial size circuits. 
Their results are based on the idea that the sender can choose a permutation and some key that is computable by the decoder but not by the channel, since it operates with low complexity. 
In some loose sense, their results are an example of our framework when the channel has bounded circuit complexity, i.e. the bounded resource is circuit complexity of the error. 

\myindent 
\cite{OPS} obtain \LDCs{} with constant information and error rates over the binary alphabet against computationally bounded errors, using a small number of queries to the corrupted word; specifically they can achieve any $\omega(\log\kappa)$ query complexity, where $\kappa$ is the desired security parameter. 
However, their results not only assume the existence of one-way functions, but also once again assume a predetermined private key known to both the encoder and decoder but not the channel, similar to~\cite{lipton_new_1994}. 
Analogous to the improvements of \cite{micali_etal_computationally_bounded_noise} for classical error codes, \cite{HemenwayO08, HemenwayOSW11} construct public-key \LDCs, assuming the existence of $\Phi$-hiding schemes \cite{CachinMS99} and IND-CPA secure cryptosystems.

\myindent 
Ben-Sasson \etal~\cite{Ben-SassonGHSV06} introduce the concept of {\em relaxed locally decodable codes} (\RLDCs) as an alternative means of decreasing the tradeoffs between rate and locality in classical \LDCs. 
In contrast to \LDCs, the decoding algorithm for \RLDCs{} is allowed to output $\bot$ sometimes to reveal that the correct value is unknown, though it is limited in the fraction of outputs in which it can output $\bot$. 
The \RLDCs{} proposed by Ben-Sasson \etal~\cite{Ben-SassonGHSV06} obtain constant query complexity and blocklength $n=k^{1+\epsilon}$. 
Subsequently, Gur \etal\cite{GurRR18} construct {\em relaxed locally correctable codes} (\RLCCs) with attractive properties but significant tradeoffs; they propose codes with constant query complexity and error rate but block length roughly quartic in the message length as well as codes with constant error rate and linear block length, but quasipolynomial ($(\log n)^{\O{\log\log n}}$) query complexity. 
These parameters are significantly better than classical locally correctable codes and their results immediately extend to \RLDCs, since the original message is embedded within the initial part of the encoding. 
However, these tradeoffs are still undesirable. 

\myindent 
Recently, Blocki \etal~\cite{rldc_comp_bounded} study \RLDCs\ and \RLCCs\ on adversarial but computationally bounded channels in an effort to reduce these tradeoffs. 
They obtain \RLDCs\ and \RLCCs\ over the binary alphabet, with constant information rate, and poly-logarithmic locality.
Moreover, their codes require no public-key or private-key cryptographic setup; the only setup assumption required is the selection of the public parameters (seed) for a collision-resistant hash function. 
\section{Preliminaries}

\subsection{ Notation}\label{sec: notation} \label{sec:notation}
\myindent
We use the notation $[n]$ to represent the set $\{1,2,\ldots,n\}$. 
For any $x, y \in \Sigma^n$, let $\HAM(x)$ denote the Hamming weight of $x$, i.e. the number of non-zero coordinates of $x$. 
Let $\HAM(x, y) = \HAM(x-y)$ denote the Hamming distance between the vectors $x$ and $y$.
All logarithms will be base $2$. 
For $n$ vectors $x_1, \ldots, x_n$, we use $\majority(x_1 \cdots x_n)$ to denote the vector that appears most frequently. If such a vector is not unique, then an arbitrary vector of highest frequency is chosen.
For any vector $x \in \Sigma^n$, let $x[i]$ be the $i$\th coordinate of $x$. 
We also let $x\circ y$ denote the concatenation of $x$ with $y$ and $x\oplus y$ denote the bitwise XOR of $x$ and $y$. For a randomized function $f(\cdot)$, the notation $f(\cdot ;R)$ will be used to denote that $f(\cdot)$ uses random coins $R$ as its randomness.
A function $\negl(\secpar)$ is said to be \emph{negligible} in \secpar if $\negl(\secpar) \in o\left( \left|\frac{1}{\mathsf{poly}(\secpar)}\right|\right)$ for any non-zero polynomial $\mathsf{poly}(\cdot)$. Finally, we distinguish between inputs and parameters to a function $f$ as follows: $f(\text{inputs}\cdots)\parameters{\text{parameters} \cdots }$.

\subsection{Locally Decodable Codes}\label{sec:Locally-Decodable-Codes}
\myindent
We consider the setting where sender \sender encodes a \emph{message} \msg into a \emph{codeword} \codeword using an \emph{encoding algorithm} so that \codeword is sent over noisy channel \channel, which then hands over the possibly corrupted codeword \corrcodeword to \receiver, who then uses a \emph{decoding algorithm} to obtain the original message. 
We denote $\msg \in \alphabet^\lenmsg$ and $\codeword \in \alphabet^\lencodeword$ where $\alphabet$ is the alphabet. 
We denote the alphabet size by $\lenalphabet = |\alphabet|$. 
We consider the model where \corrcodeword corresponds to \codeword with some symbols replaced with others in \alphabet. 
The term \emph{corruptions} refers to such symbol replacements within \codeword, with a single corruption meaning a single symbol replacement, so that $\corrcodeword \in \alphabet^\lencodeword$. 
The encoding and decoding algorithms are denoted by $\encoder: \alphabet^\lenmsg \rightarrow \alphabet^\lencodeword$ and $\decoder:\alphabet^\lencodeword \rightarrow \alphabet^\lenmsg$. 
We use the terms sender, encoder, and encoding algorithm interchangeably, and similarly for receiver, decoder, and decoding algorithm. 

\myindent
A \emph{code} is an encoder-decoder pair. 
The \emph{information rate} or simply \emph{rate} of the code is the ratio $\lenmsg/\lencodeword$, so that a lower rate corresponds to a larger amount of information redundancy introduced by the code. 
The message length, codeword length, and alphabet size characterize a \emph{coding scheme}.
Coding schemes with high rate and low alphabet size are desired.

\myindent
An error correcting code allows the decoder to recover the entire original message \msg by reading the entire \corrcodeword. 
It is also possible to construct codes that only need to read a few symbols of \corrcodeword rather than the entire message to recover a small part of the message. 
Such codes are called \emph{locally decodable codes} (\LDC), and will be the focus of this work. 
An \LDC has \emph{locality} \loc, \emph{error rate} \errorrate and \emph{error correction probability} \probrecov if any character of \msg may be recovered with probability at least \probrecov by making at most \loc queries to \corrcodeword, even when the channel corrupts \errorrate fraction of all symbols of \codeword to generate \corrcodeword. 
We use the terms \emph{query complexity} and locality interchangeably. 
When $\errorrate$ and $\probrecov$ are clear from context (as constants), the scheme may be referred to as an \emph{$\loc$-\LDC}. 
Naturally, \LDCs\ with low locality, high error rate, and high error correction probability are desired.

\subsection{Definitions}\label{sec:definitions}
The focus of this work will be the construction of \LDCs{} (Section \ref{sec:model}) for \emph{resource-bounded} channels (Section \ref{sec:pROM}). In this section, we present several building blocks that we will require in our constructions ---  \emph{\LDCsvariant}, \emph{private-\LDCs} and \emph{safe functions}. We first give two classical definitions pertaining to \LDCs{} that compactly summarize our discussion in Section \ref{sec:Locally-Decodable-Codes}. 

\begin{definition}
A \emph{$(\lencodeword,\lenmsg)_\lenalphabet$-coding scheme} $C\parameters{K,k,q} = (\Enc,\Dec)$ is a pair of encoding $\Enc:\alphabet^\lenmsg \rightarrow \alphabet^\lencodeword$ and decoding $\Dec : \alphabet^\lencodeword \rightarrow \alphabet^\lenmsg$ algorithms where $|\alphabet| = \lenalphabet$. The \emph{information rate} of the scheme is defined as $\frac \lenmsg \lencodeword$.
\end{definition}

\begin{definition}
\label{def:ldc}
A $(\lencodeword,\lenmsg)_\lenalphabet$-coding scheme $C\parameters{K,k,q} = (\Enc,\Dec)$ is an \emph{$(\loc, \errorrate, \probrecov)$-locally decodable code (\LDC)} if \Dec, with query access to a word $\corrcodeword$ such that $\HAM(\Enc(\msg),\corrcodeword) \leq \errorrate\lencodeword$, on input index $i \in [k]$, makes at most \loc queries to \corrcodeword and outputs $\msg_i$ with probability at least $p$ over the randomness of the decoder.
\end{definition}

Next, we present a simple variant of \LDCs{} which we denote by \LDCsvariant. These will be very similar to  \LDCs{} except that they are required to decode the entire original message while making as few queries to the corrupted codeword as possible. They are defined with respect to the same setting as in Section \ref{sec:Locally-Decodable-Codes}. 
\begin{definition}
\label{def:ldcvariant}
A $(\lencodeword,\lenmsg)_\lenalphabet$-coding scheme $C\parameters{K,k,q} = (\Enc,\Dec)$ is an \emph{$(\loc, \errorrate, \probrecov)$-\LDCvariant} if \Dec, with query access to a word $\corrcodeword$ such that $\HAM(\Enc(\msg),\corrcodeword) \leq \errorrate\lencodeword$, makes at most \loc queries to \corrcodeword and outputs $\msg$ with probability at least $p$ over the randomness of the decoder.
\end{definition}

We remark that it will be typically desired that for an \LDCvariant{} $\code\parameters{\lencodeword,\lenmsg,\lenalphabet}$, the locality be $\O{\lenmsg}$ even when $\lencodeword$ is very large. We now move on to define private-\LDCs{} analogous to Definition \ref{def:ldc} as an alternative to that given by \cite{OPS} -- refer to Appendix \ref{appendix:privateldc} for an overview of \cite{OPS}.


\begin{figure}[!htb]
\footnotesize
\begin{mdframed}
$\mathtt{{priv-LDC-Sec-Game}}\parameters{\attacker, \msg, \secpar, \errorrate, p}:$
\begin{enumerate}
\item The challenger generates a secret key $\sk \leftarrow \mathsf{GenKey}(1^\secpar)$, computes the codeword $\codeword \leftarrow \Enc(\msg, \secpar, \sk)$ for the message $\msg$ and sends the codeword $\codeword$ to the attacker.
\item The attacker outputs a corrupted codeword $\codeword'\leftarrow  \attacker\left(x,  \codeword, \secpar, \errorrate,p, \lenmsg, \lencodeword \right) $ where $\codeword' \in {\alphabet}^{\lencodeword}$ should have hamming distance at most $\errorrate \lencodeword$ from $\codeword$. 

\item
The output of the experiment is determined as follows:
    $$\mathtt{priv-LDC-Sec-Game\parameters{\attacker, x,  \secpar, \errorrate, p}} = \begin{cases}
        1& \text{if $\HAM(\codeword,\codeword') \leq \errorrate \lencodeword$ and $\exists i \leq \lenmsg$ s.t. $\Pr[\Dec^{\codeword'}(i,\secpar, \sk) = \msg_i] < p$} \\
        0& \text{otherwise}\\
    \end{cases}$$
    If the output of the experiment is $1$ (resp. $0$), the attacker $\attacker$ is said to \emph{win} (resp. \emph{lose}) against $\mathsf{C}$.
\end{enumerate}
\end{mdframed}
\vspace{-0.1in}
\caption{$\mathtt{{priv-LDC-Sec-Game}}$ defining the interaction between an attacker and an honest party 
}
\label{fig:priv-ldc-sec-game}
\end{figure}

\begin{definition}\label{def:game-based-private-LDC} (One-Time Private Key LDC)
A triplet of probabilistic algorithms $\mathsf{C}\parameters{\lencodeword,\lenmsg,\secpar} = (\mathsf{GenKey}, \Enc, \Dec)$ is an \emph{$(\loc, \errorrate, \probrecov, \epsilon, \class)$-private locally decodable code (private \LDC)} against a class $\class$ if $\Dec$ makes at most \loc queries and for all attackers $\attacker \in \class$ and all messages $\msg \in \alphabet^\lenmsg$ we have 
$$\Pr[\texttt{priv-LDC-Sec-Game}\parameters{\attacker, \msg,  \secpar, \rho, p} = 1] \leq \epsilon$$
where the probability is taken over all the random coins of \attacker{} and $\mathsf{GenKey}$. If $\class$ is the set of all (computationally unbounded) attackers we simply say that the scheme is a \emph{$(\loc, \errorrate, \probrecov, \epsilon)$-private LDC}.

\end{definition}

\myindent Our contributions in the subsequent sections will assume that the coding scheme and channel all have access to a random oracle. Furthermore, we assume that the channel is a \pROM algorithm with respect to this random oracle (refer to the initial discussion in Section \ref{sec:pROM} for an overview of the \pROM model). The following definition establishes a notion of \emph{privacy} against classes (i.e. sets) of adversarial channels in terms of ``hard to compute'' functions. 


\begin{definition}[Safe Function]\label{def:safe_function}
We say that a function $f:\{0,1\}^n \rightarrow \{0,1\}^*$ is \emph{$\delta$-safe} for a class $\class$ of algorithms if for all $\attacker \in \class$ we have \[\Pr\big[\attacker(x) = f(x)\big] \leq \delta\] 
where the probability is taken over the random coins of \attacker{} and the selection of an input $x \in \{0,1\}^n$. 
If the function $f=f^{\oracleH}$ is defined using a random oracle, then the probability $\Pr\big[\roattacker(x) = f^{\oracleH}(x)\big]$ is also taken over the selection of the random oracle $\oracleH$.
\end{definition}
We will use the notation \safe{} to denote a $\delta-$safe function for class $\class$. In the above definition, we usually think of $\delta$ as being a negligibly small parameter. 
We remark that in the parallel random oracle model, one can construct functions with sharp thresholds on the required resources. 
For example, the function $\mathsf{H}^{t+1}(x)$ is trivial to compute using at most $t+1$ sequential queries to $\mathsf{H}:\{0,1\}^*\rightarrow \{0,1\}^2$, but {\em any} parallel algorithm making at most $q$ queries over $t$ rounds succeeds with probability at most $\delta = (t^2 + tq)/2^w$. 

\paragraph{Precomputation.} 
Definition \ref{def:safe_function} can be extended to consider an attacker who is allowed to perform precomputation with the random oracle $\oracleH$ before receiving the input $x$. In particular, we could consider a pair of oracle algorithms $(\attacker_1,\attacker_2)$ where $\attacker_1^{\oracleH}(m)$ outputs an $m$-bit hint $\sigma \in \{0,1\}^m$ for $\attacker_2$ after making at most $q$ queries to $\oracleH$. We could modify the definition to require that for all $\attacker_2 \in \class$ we have
\[\Pr\bigg[\attacker_2^{\oracleH}(x,\attacker_1^{\oracleH}(m)) = f^{\oracleH}(x)\bigg] \leq \delta \ , \] 
where the randomness is taken over the selection of $x$, the random oracle $\oracleH$, and the random coins of $\attacker_2$. Here, $\attacker_1^{\oracleH}(m)$ (precomputation) is not necessarily constrained to be in the same class $\class$ as $\attacker_2$. 

\myindent
We remark that for $k=m/w$, a precomputing attacker can succeed with probability at least $ k/2^n$ by having $\attacker_1^{\oracleH}(m)$ output the hint $\sigma = f^{\oracleH}(1),\ldots, f^{\oracleH}(k)$.  Then $\attacker_2^{\oracleH}(x,\sigma)$ first checks if $x \in \{1,\ldots, k\}$ and, if so,  simply returns the  output $f^{\oracleH}(x)$ which is already recorded in the hint $\sigma$. 
Thus, we need the length $n$ of the random nonce $x$ to be sufficiently large to resist brute-force precomputation attacks. 
By contrast, if the attacker does not get to perform any precomputation then $\delta$ can be negligible even when $n=\O{1}$. 

\myindent
All of the safe functions we consider would also be secure under this stronger notion. 
For example, $\mathsf{H}^{t+1}(x)$ is $\delta$-safe for $\delta = \O{ (\numOracleQueries t+t^2)/2^w + \numOracleQueries t/2^n}$ where $x$ is a random $n$ bit string, $\attacker_1$ makes at most $\numOracleQueries$ total random oracle queries, and $\attacker_2$ makes at most $\numOracleQueries$ total queries in at most $t$ rounds to $\oracleH$. 
In our \LDC{} constructions we select a random nonce of length $\Omega(\log^{1+\eps} \kappa)$ to ensure that a precomputing attacker fails. 

\subsection{Our Model} \label{sec:model}

\begin{figure}[!htb]
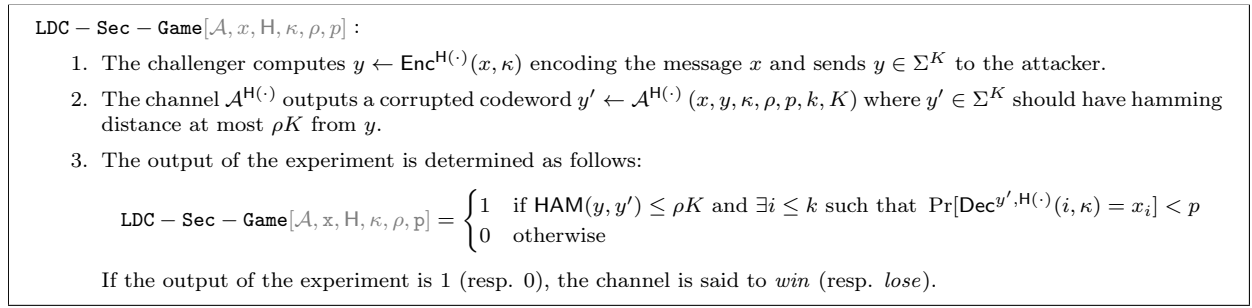

\footnotesize
\begin{mdframed}
$\mathtt{{LDC-Sec-Game}}\parameters{\attacker, \msg, \mathsf{H}, \secpar, \errorrate, p}:$
\begin{enumerate}

\item
The challenger computes $\codeword \leftarrow \Enc^{\oracleH}(\msg, \secpar) $ encoding the message $\msg$ and sends $\codeword \in \alphabet^\lencodeword$ to the attacker.
\item
The channel $\roattacker$ outputs a corrupted codeword $\codeword'\leftarrow  \roattacker\left(\msg,  \codeword, \secpar, \errorrate,p, \lenmsg, \lencodeword \right) $ where $\codeword' \in {\alphabet}^{\lencodeword}$ should have hamming distance at most $\errorrate \lencodeword$ from $\codeword$. 
\item
The output of the experiment is determined as follows:
    $$\mathtt{LDC-Sec-Game\parameters{\attacker, \msg, \mathsf{H}, \secpar, \errorrate, p}} = \begin{cases}
        1& \text{if $\HAM(\codeword,\corrcodeword) \leq \errorrate \lencodeword$ and $\exists i \leq \lenmsg$ such that } \Pr[\Dec^{\codeword',\oracleH}(i, \secpar) = \msg_i] < p\\
        0& \text{otherwise}\\
    \end{cases}$$
    If the output of the experiment is $1$ (resp. $0$), the channel  is said to \emph{win} (resp. \emph{lose}).
\end{enumerate}
\end{mdframed}
\vspace{-0.1in}
\caption{$\mathtt{{LDC-Sec-Game}}$ defining the interaction between an attacker and an honest party 
}
\label{fig:ldc-sec-game}
\end{figure}

\myindent 
We first define an experiment to model the interaction between a code and an algorithm from a class of \pROM algorithms adversarial against the code. For random oracle \oracleH, let $\mathsf{C} = (\Enc^{\oracleH},\Dec^{\oracleH})$ be a $(\lencodeword,\lenmsg)_\lenalphabet$-coding scheme in the random oracle model and let $\class$ be a class of \pROM algorithms. Then, the interaction of $\roattacker \in \class$ having error rate $\errorrate$, with the code $\mathsf{C}$ is defined in Figure~\ref{fig:ldc-sec-game} (analogous to {\texttt{priv-LDC-Sec-Game}} defined in Figure~\ref{fig:priv-ldc-sec-game}). Here, the security parameter $\secpar$, and the decoding probability $p$ are also given as inputs to the game.
We now formally define a notion of  \LDCs{} analogous to Definition \ref{def:ldc}, but with respect to general classes of adversarial (\pROM) channels.


\begin{definition}
Let \class be a class of \pROM algorithms. A $(\lencodeword,\lenmsg)_\lenalphabet$-coding scheme $\mathsf{C}\parameters{K,k,q} = (\Enc^{\oracleH},\Dec^{\oracleH})$ is an \emph{$(\loc, \errorrate, \probrecov, \epsilon, \class)$-locally decodable code (\LDC)} if $\Dec^{\oracleH}$ makes at most \loc queries and for all $\roattacker \in \class$ and all messages $\msg \in \alphabet^{\lenmsg}$ we have
$$\Pr[\mathtt{LDC-Sec-Game\parameters{\attacker,\msg, \mathsf{H}, \secpar, \errorrate, p} = 1}] \leq \epsilon$$
where the probability is taken over the random coins of \roattacker and the selection of the random oracle $\mathsf{H}$.
\end{definition}
\myindent We remark that our codes need not require that each message have the same length. 
Jumping slightly ahead, longer messages only need proportionally longer repetition codes to guarantee transmittance of the secret key. 
However for the sake of presentation, we use notation for fixed length messages.
\section{Constructions}
\myindent  
We begin by discussing the use of safe functions in Section \ref{sec:safe_function} and give several examples of constructing such functions in Section \ref{sec:safe}. 
We then show how allowing an encoder/decoder pair with enough resources to compute safe functions can effectively generate a random shared secret key between the pair. 
This secret key can then be bootstrapped into existing private \LDC\ constructions to give codes against resource bounded adversaries. 
We give our final framework in Section \ref{sec:framework} and the main proofs in Sections \ref{sec:construction_properties} and \ref{sec:securityAndDecodingProb}.

\subsection{Using Safe Functions}\label{sec:safe_function}
\myindent 
Let \class be a class of algorithms with safe function \safe. For some input $x \in \{0,1\}^n$ to $\roattacker\in\class$, we will be interested in bounding the probability of the undesirable event where the \roattacker queries the random oracle at any string of the form $y \circ  \safe(x)$ with $y \in \{0,1\}^{\lceil \log_2 \alpha \rceil}$. 
In the absence of such an event, $\mathsf{H}(\safe(x))$ would information theoretically appear random to \roattacker. 
Lemma \ref{lem:prob_bad_event_attacker} shows that such an event may only happen with negligible probability $\numOracleQueries\epsilon$ where $\numOracleQueries$ is the total number of random oracle queries.

\begin{lemma}\label{lem:prob_bad_event_attacker}
For a some class \class of \pROM algorithms with $\delta-$safe function $\safe \zo^n\rightarrow\zo^*$, let $\bad_{\attacker}$ be the event that on some input $x \in \{0,1\}^n$, $\roattacker \in \class$ queries the random oracle at $\alpha \circ \safe(x)$ for any $\alpha > 0$. Then $\Pr[\bad_{\attacker}] \leq \numOracleQueries\delta$, where $\numOracleQueries$ is the number of oracle queries made by \roattacker.
\end{lemma}
\begin{proof}
We prove the claim by a reduction argument. 
By way of contradiction, suppose there exists a $\mathcal{B}^{\oracleH} \in \class$ such that on input string $x$, $\mathcal{B}^{\oracleH}$ makes $\numOracleQueries$ queries to the random oracle $\oracleH$ and $\Pr[\bad_{\mathcal{B}}] > \numOracleQueries\epsilon$. 
We construct an adversary $\roattacker$ as follows: on input $x$, the adversary
\begin{itemize}
    \item Simulates $\mathcal{B}^{\oracleH}$ with input $x$
    \item Keeps track of all $\numOracleQueries$ queries by which $\mathcal{B}^{\oracleH}$ queries the random oracle
    \item On termination of $\mathcal{B}^{\oracleH}$, returns the suffix of length |\safe(x)| from one of the $\numOracleQueries$ queries selected uniformly at random
\end{itemize}
However, we know that $\mathcal{B}^{\oracleH}$ queries the random oracle at $\alpha \circ \safe(x)$ with probability $> \numOracleQueries\delta$. Since $\roattacker$ picks one of $\mathcal{B}^{\oracleH}$'s queries at random, $\Pr[\roattacker(x) = \safe(x)] > \delta$, which contradicts the definition of $\delta-$safe function.
\end{proof}

\myindent
Assuming that \roattacker never queries the random oracle at any point of the form $y \circ  \safe(x)$ with $y \in \{0,1\}^{\lceil \log_2 \alpha \rceil}$ (for some $\alpha > 0$) we can view each $\mathsf{H}(y \circ  \safe(x))$ as a fresh $w$-bit string. 
Thus, we can obtain a random $w\alpha$-bit string by concatenating all of the labels $\mathsf{H}(y \circ  \safe(x))$ for each $y \in \{0,1\}^{\lceil \log_2 \alpha \rceil}$. This motivates the following definition of an \emph{expansion family} which will be used in subsequent sections. 

\begin{definition}[Expansion Family]\label{def:expansion-function}
For random oracle \oracleH\ the expansion family of functions 
$\{\mathsf{E}^{\oracleH}_{\alpha}\}_{\alpha = 1}^\infty$ where each function $\mathsf{E}^{\oracleH}_{\alpha}:\{0,1\}^* \rightarrow \{0,1\}^{\alpha w}$ is defined as $\mathsf{E}^{\oracleH}_{\alpha}(x) = \mathsf{H}(1 \circ x) \circ \mathsf{H}(2 \circ x) \circ \cdots \circ \mathsf{H}(\alpha \circ x)$, where the prefix $i \in [\alpha]$ of $x$ for each oracle query in the definition is expressed in binary using $\lceil\log_2\alpha\rceil$ bits.
\end{definition}

\subsection{Framework for \LDCs\ against Resource Bounded Channels}\label{sec:framework}

Our aim in this section will be to achieve \LDCs\ having no asymptotic loss in rate, query complexity, or success probability of private locally decodable codes. In contrast to the private \LDC setting, we will assume no private (or public) key setup assumptions. We will also aim for \LDCs\ that may be used for multiple (polynomial) rounds of communication, a notion which we describe later in the section.\\

\myindent Let $\codeLDC\parameters{\lencodewordLDC,\lenmsgLDC} = (\encLDC, \decLDC)$ be an $(\locLDC, \errorrateLDC, \probrecovLDC)-$\LDCvariant (recall Definition \ref{def:ldcvariant}). 
Furthermore, let $\codePriv\parameters{\lencodewordPriv,\lenmsgPriv} = (\encPriv, \decPriv, \genkeyPriv)$ be a $(\locPriv, \errorratePriv, \probrecovPriv, \epsilonPriv)-$private \LDC (recall Definition \ref{def:game-based-private-LDC}). Against classes of \pROM algorithms permitting $\delta-$safe functions, our encoder will use \codeLDC to bootstrap off of \codePriv even in the absence of shared private randomness with the decoder.

\paragraph{Framework Overview:}

The encoding algorithm first samples a random seed \controlinfo of modest length (\lenmsgLDC). By embedding an encoding of \controlinfo (via \codeLDC) in our final codeword, we can ensure that our decoder will also have access to \controlinfo.
Let the channel, over which the communication happens, belong to a class \class of \pROM algorithms (w.r.t. random oracle \oracleH) permitting some $\delta$-safe function $\safe : \{0,1\}^{\lenmsgLDC} \rightarrow \{0,1\}^{*}$. Even though the channel has access to the seed \controlinfo, it will be unable to compute $\safe(\controlinfo)$ by definition of the safe function. Thus $\randOracleH{\safe(\controlinfo)}$ is effectively a random string to the channel. We can expand this randomness via an expansion function (Definition \ref{def:expansion-function}), and use \genkeyPriv with this randomness to compute a key. The computed key is effectively secret from the channel and can be used in conjunction with \encPriv to obtain an encoding of any input message. Note that since the decoder also has access to \controlinfo, it may also compute the secret key using exactly the same procedure and use this key in conjunction with \decPriv to perform the required decoding. Thus the use of \codeLDC, safe and expansion functions on a random seed reduces the setting to that of \codePriv. Our framework is parameterized by \parameters{\safe, \codeLDC, \codePriv}.




\begin{figure}
  \centering
  \includegraphics[width=0.9\linewidth]{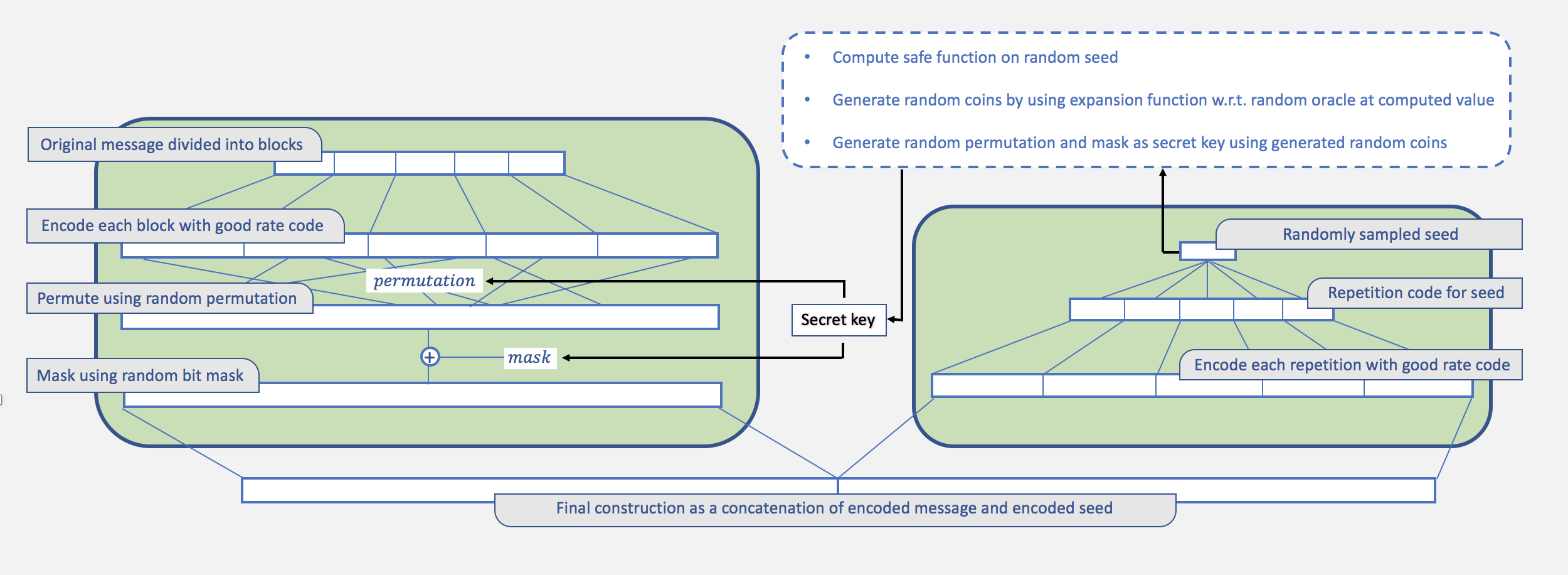}
  \caption{Instantiation of framework for \LDCs\ against adversaries permitting safe functions}
  \label{fig:framework_LDCs}
\end{figure}

\paragraph{Explicit Constructions:}
We provide explicit constructions of \LDCs{} against adversarial \pROM{} channels permitting $\delta-$safe functions by instantiating the framework discussed above. Figure \ref{fig:framework_LDCs} gives an overview of the instantiation. For private \LDCs{}, we will make use of the constructions of Theorem \ref{thm:ops-priv-loc-dec-code}. Furthermore, we instantiate \codeLDC as follows: The encoder encodes the seed with a standard constant rate error correcting code -- we instantiate this with Justesen codes -- composed with a repetition code. The local decoder then randomly samples seed-encodings and takes a majority vote over the decoded samples to determine the seed. We refer the reader to Appendix \ref{appendix:justesen} for a formal explanation of this \LDCvariant{} instantiation.\\

Detailed descriptions of our encoder (\Encro) and decoder (\Decro), given a message $x$, security parameter \secpar, and random oracle \oracleH, may be described in Figure~\ref{fig:ldc}. In particular, our framework lead to the following theorem.

\begin{figure}[!htb]
\begin{center}
\footnotesize
\begin{tabu} to \textwidth {| X | X |}
    \hline
    &\\
    $\mathbf{\Encro(x, \secpar)}\parameters{ \safe, \codeLDC,\codeFinal}:$
\begin{enumerate}
    \item Sample a random seed of length \lenmsgLDC. 
    
    $\controlinfo \leftarrow \{0,1\}^{\lenmsgLDC}$
    
    \item Encode random seed using an \LDC. 
    
    $\codewordLDC := \encLDC(\controlinfo)$
    
    \item Generate randomness uncomputable by channel via safe and expansion functions.
    
    $R := \mathsf{E}^{\oracleH}_\tau(\safe(\controlinfo))$
    
    \item Generate a secret key from the randomness.
    
    $\skFinal := \genkeyPriv(\secpar; R)$
    \item Use private \LDC encoder with generated key.
    
    $\codewordPriv :=$ $\encPriv(x, \secpar, \skFinal)$
    \item \textbf{Output} $\codewordPriv \circ \codewordLDC$
\end{enumerate}
& 
$\mathbf{\Dec^{\oracleH, \codewordPriv' \circ \codewordLDC'}_{\mathsf{final}}(i, \secpar)} \parameters{ \safe, \codeLDC,\codeFinal}:$
\begin{enumerate}
    \item Decode the original random seed.
    
    $\controlinfo := \decLDC^{\codewordLDC'}()$
    
    \item Compute randomness used by encoder.
    
    $R := \mathsf{E}^{\oracleH}_\tau(\safe(\controlinfo))$
    \item Compute secret key used by encoder.
    
    $\skFinal := \genkey_\mathsf{OPS}(\secpar; R)$
    \item Use private \LDC decoder with computed key.
    
    \textbf{Output} $\decPriv^{\codewordPriv'}(i, \skFinal)$
\end{enumerate} \\
    &\\
    \hline
\end{tabu}
\end{center}
\vskip -0.1in
\caption{Encoding and decoding algorithms for our \LDC{} construction.}
\label{fig:ldc}
\end{figure}

\begin{theorem}\label{thm:framework}
 Let $\codePriv\parameters{\lencodewordPriv,\lenmsgPriv} = (\encPriv, \decPriv, \genkeyPriv)$ be a $(\locPriv, \errorratePriv, \probrecovPriv, \epsilonPriv)-$private \LDC and $\codeLDC\parameters{\lencodewordLDC,\lenmsgLDC, \secpar} = (\encLDC, \decLDC)$ be an $(\locLDC, \errorrateLDC, \probrecovLDC)-$\LDCvariant. Then for any class \class of \pROM algorithms admitting a $\delta-$safe function $\safe: \{0,1\}^{\lenmsgLDC} \rightarrow \{0,1\}^{*}$, the $(\lencodewordFinal, \lenmsgFinal)_2$ coding scheme in the random oracle model $\codeFinal\parameters{ \safe, \codeLDC, \codePriv} = (\encFinal,\decFinal)$ is an $(\locFinal, \errorrateFinal,\probrecovFinal,\epsilonFinal)$-\LDC with $\lenmsgFinal = \lenmsgPriv$, $\lencodewordFinal = \lencodewordLDC + \lencodewordPriv$, $\locFinal = \locFinal + \locLDC$, $\errorrateFinal = \frac{1}{\lencodewordLDC + \lencodewordPriv}\min\{\errorrateLDC\lencodewordLDC, \errorratePriv\lencodewordPriv\}$, $\probrecovFinal \geq 1 -  \lenmsgPriv(2 - \probrecovPriv - \probrecovLDC)$, $\epsilonFinal \leq \epsilonPriv + \numOracleQueries\delta$. Here $\numOracleQueries$ is an upper bound on the number of queries any algorithm $\roattacker \in \class$ makes to the random oracle \oracleH.

\end{theorem}

\myindent The final codeword generated by \encFinal is simply the concatenation of the codewords generated by \encPriv and \encLDC, resulting in $\lencodewordFinal = \lencodewordLDC+\lencodewordPriv$. By construction, the only queries \decFinal makes to the corrupted codeword are during the executions of \decLDC and \decPriv. This gives the locality $\locFinal = \locLDC + \locPriv$. Furthermore for correct overall decoding, it is necessary that the individual codes are correctly decoded. Thus the total errors that the code can tolerate is bounded by the maximum number of errors any individual one of the codes can tolerate. This gives the claimed (worst case) error rate. We emphasize that the proofs of the bounds on the decoder's success probability and the security of the framework is much more involved than the above discussion and is included in Section \ref{sec:construction_properties} and \ref{sec:securityAndDecodingProb}.
In particular, we show that no adversary admitting $\delta$-safe functions can distinguish between the encodings of \encFinal and those of \encPriv with random strings appended to them. Furthermore, even the decoder, who has no computational restrictions and gets the appropriate secret key used during the respective encoding processes may not make this distinction, thereby effectively reducing the security of \codeFinal to that of \codePriv with negligible loss. The following two corollaries exhibit decoding probability vs locality tradeoffs when our framework is instantiated with the \LDCsvariant{} in Appendix \ref{appendix:justesen} and the private-\LDCs{} of Appendix \ref{appendix:privateldc}.

\begin{corollary}\label{cor:p_constant_main}
For security parameter \secpar, a class \class of \pROM adversaries admitting $\delta-$safe function $\safe: \{0,1\}^{\log^{1+\eps}{\secpar}} \rightarrow \{0,1\}^{*}$ where $\eps > 0$ and for every $k > 0$ such that $k = \poly(\secpar)$ where $\poly$ is any non-zero polynomial, there exists a $(\beta k,k)_2$ coding scheme in the random oracle model that is an $(\loc, \errorrate, \probrecov, \epsilon, \class)-$\LDC where $\loc = (\alpha + 1)\log^{1+\eps}\secpar$ (such that $\alpha \geq 17$), $\errorrate$ is a constant, $p$ is a constant dependent on $\alpha$, and $\epsilon \leq \negl(\secpar) +\numOracleQueries\delta$. Here $\beta$ is a constant, $\negl(\secpar)$ is a negligible function of \secpar and $\numOracleQueries$ is an upper bound on the total queries any algorithm in $\class$ makes to the random oracle.
\end{corollary}

\begin{corollary}\label{cor:p_whp_main}
For security parameter \secpar, a class \class of \pROM adversaries admitting $\delta-$safe function $\safe: \{0,1\}^{\log^{1+\eps}{\secpar}} \rightarrow \{0,1\}^{*}$ where $\eps > 0$ and for every $k > 0$ such that $k = \poly(\secpar)$ where $\poly$ is any non-zero polynomial, there exists a $(\beta k,k)_2$ coding scheme in the random oracle model that is an $(\loc, \errorrate, \probrecov, \epsilon, \class)-$\LDC where $\loc = (1 + 24\log^{1+\eps}{\secpar}) \log^{1+\eps}{\secpar}$, $\errorrate$ is a constant, $p \geq (1 - \negl_1(\secpar))$, and $\epsilon \leq \negl_2(\secpar) +\numOracleQueries\delta$. Here $\beta$ is a constant, $\negl_1(\secpar)$ and $\negl_2(\secpar)$ are negligible functions of \secpar, and $\numOracleQueries$ is an upper bound on the total queries any algorithm in $\class$ makes to the random oracle.
\end{corollary}

\paragraph{Precomputation.} We remark that Steps $1$-$4$ of \encFinal may be precomputed. This may be advantageous in some settings to speed up encoding time as the sender may precompute multiple $(\finalsk, \jrepcontrolcode)$ pairs. When a message is ready to be encoded, the sender then simply needs to generate $\opscode$ using $\finalsk$ and append \jrepcontrolcode\ to generate the final codeword. However, we do note that this precomputation must be done after the selection of the random oracle, and that such precomputation is not possible for \Decro.

\paragraph{Multi-round Communication.} 
Existing constructions of private \LDCs~\cite{OPS} are secure only for a single \emph{round} of communication (see Appendix \ref{appendix:privateldc} for details on the round-based game between the encoder/decoder and the channel in the private \LDC\ setting). 
We may generalize our model to be in terms of rounds as well, where each round runs an instance of the experiment $\mathtt{{LDC-Sec-Game}}$ defined in Section \ref{sec:model}. 
We remark that our codes work for this generalized model as well. In every round of the experiment, the encoder can sample a fresh random seed $\controlinfo$. 
This is not directly possible in the existing private \LDC constructions  as an attacker listening to the decoder's queries may learn information about the secret key after a single round of communication. 
For this Ostrovsky \etal introduce a new construction which hides the secret key behind a layer of encryption, which in turn increases the locality of their final constructions to $\omega(\log^2\kappa)$. 


\subsection{Two-Phase Hybrid Distinguisher Argument} \label{sec:construction_properties}

To prove the security of the \LDC{} framework in section \ref{sec:framework}, our approach is to argue the following: if any channel wins the \texttt{LDC-Sec-Game} against an instantiation of our \LDC constructions (\Encro, \Decro), then this channel can win the \texttt{priv-LDC-Sec-Game} against its constituent private-\LDC (contradicting its security guarantee). \\

\noindent
\textbf{Standard Hybrid Argument Failure:} 
A natural attempt to prove this, yet one that fails, is to use the following standard hybrid argument. In the first hybrid we use our original encoding scheme \encFinal to obtain a codeword $\privcodewordHybridZero \circ \ldccodewordHybridZero$. In the second hybrid, we replace the second component with an encoding of a random unrelated nonce to get $\privcodewordHybridOne \circ \ldccodewordHybridOne$. Here $\ldccodewordHybridOne$ is an encoding of some random nonce which is sampled completely independent of the message encoding $\privcodewordHybridOne$. We would like to argue that the two hybrids are indistinguishable and conclude that a resource bounded channel cannot fool the local decoder from original encoding scheme (first hybrid) --- since we cannot fool the private-\LDC local decoder in the second hybrid. However,  if the distinguisher $\distinguisher$ is able to evaluate the safe-function then the hybrids are trivially distinguishably. On the other hand, if we assume that the distinguisher $\distinguisher$ is resource bounded like the channel then indistinguishability does not suffice to argue that the local decoder i.e., fooling the decoder does not yield a resource bounded distinguisher $\distinguisher$  since the decoder is not constrained in the same way as the resource bounded channel. \\

\noindent
\textbf{Two-Phase Argument Overview:}
We address the previous issue by introducing a \emph{two-phase distinguisher game} defined over adversary/distinguisher pairs. In the first phase of this game, a random coin toss $b \in \{0,1\}$ randomly selects one of the hybrid encoders to encode a message. The selected hybrid hands its encoding $\privcodewordHybridExp \circ \ldccodewordHybridExp$ to the adversary \roattacker{} which outputs a corrupted codeword $\corrcodewordHybridExp$.
In the second phase, the distinguisher $\distinguisher$ is given the initial message \msg, the corrupted codeword $\corrcodewordHybridExp$, along with the secret key $\skHybridExp$ used to obtain $\privcodewordHybridExp$, and tries to predict the value of $b$, i.e., which hybrid encoder was used. 
An important point to note is that $\distinguisher$ is not constrained in any way. However, it is not given access to the random oracle. 
We show (Lemma \ref{lem:advAD_upper_bound}) that for any such attacker-distinguisher pair, the distinguisher succeeds at guessing which hybrid encoding was used with at most negligible probability.
The two phase hybrid argument allows us to reason about our original goal: the probability that the channel fools the honest decoder. In particular, a channel that wins the \texttt{LDC-Sec-Game} with non-negligible probability can be used in phase 1 in conjunction with a distinguisher that can simulate the decoding algorithm (with the correct key) in phase 2 to distinguish between the hybrids with non-negligible probability. This gives the required contradiction (Lemma \ref{lem:distinguisher_contradiction}). We formally define the two hybrid encoders in Figure~\ref{fig:hybrids}.\\

\begin{figure}[!htb]
\begin{center}
\footnotesize
\begin{tabu} to \textwidth {| X | X |}
    \hline
    &\\
    $\mathbf{\encHybridZero(\msg, \secpar)}$\parameters{\safe, \codeLDC, \codePriv}: (\textbf{same as Figure \ref{fig:framework_LDCs}})
    
    \begin{enumerate}
    
     \item Sample a random seed of length \lenmsgLDC. 
    
    $\seedHybridZero \leftarrow \{0,1\}^{\lenmsgLDC}$
    
    \item Encode random seed using an \LDCvariant. 
    
    $\ldccodewordHybridZero := \encLDC(\seedHybridZero)$
    
    \item Generate randomness uncomputable by channel via safe and expansion functions.
    
    $\randomnessHybridZero := \mathsf{E}^{\oracleH}_\tau(\safe(\seedHybridZero))$
    
    \item Generate a secret key from the randomness.
    
    $\skHybridZero := \genkeyPriv(\secpar; \randomnessHybridZero)$
    \item Use private \LDC encoder with generated key.
    
    $\privcodewordHybridZero :=$ $\encPriv(x, \secpar, \skHybridZero)$
    \item \textbf{Output} $\privcodewordHybridZero \circ \ldccodewordHybridZero$
    
    \end{enumerate}
    
& $\mathbf{\encHybridOne(\msg, \skHybridOne, \secpar)}$\parameters{ \safe, \codeLDC, \codePriv}:

    \begin{enumerate}
    
     \item Sample a random seed of length \lenmsgLDC. 
    
    $\seedHybridOne \leftarrow \{0,1\}^{\lenmsgLDC}$
    
    \item Encode random seed using an \LDCvariant. 
    
     $\ldccodewordHybridOne := \encLDC(\seedHybridOne)$
    \item Use private \LDC encoder with input key.
    
    $\privcodewordHybridOne :=$ $\encPriv(x, \secpar, \skHybridOne)$
    \item \textbf{Output} $\privcodewordHybridOne \circ \ldccodewordHybridOne$
\end{enumerate} \\
    &\\
    \hline
\end{tabu}\\
\end{center}
\vskip -0.1in
\caption{Hybrid encoding algorithms. By design, \hybridusenc{} is the same as our proposed \LDC{} construction.}
\label{fig:hybrids}
\end{figure}

\myindent Let \roattacker{} be an adversarial channel belonging to a class \class of \pROM algorithms w.r.t random oracle \oracleH{} permitting $\delta-$safe functions. Furthermore, let $\distinguisher: \big(\zo^{*}\big)^4 \rightarrow \zo$ be a computationally unbounded algorithm. We will term \roattacker{} and \distinguisher as \emph{attacker} and \emph{distinguisher} respectively. Using the \emph{hybrid} encoders in Figure \ref{fig:hybrids}, we define the \emph{indistinguishability experiment} \indistinguishExp over all \emph{attacker-distinguisher pairs} \ADpair. Note that in this experiment, \distinguisher is provided with the secret key that the selected hybrid used during encoding, and does not have access to the random oracle. With respect to this experiment, we define the \emph{advantage} of the attacker-distinguisher pair as follows:


$$\advantageAD := \max_x \bigg|\Pr[\mathtt{Exp}_{\attacker,\distinguisher,\mathsf{H},\secpar, x} = 1] - \frac{1}{2} \bigg|$$
where the probability is taken over the randomness of $\distinguisher$, $\roattacker$, and the selection of the random oracle \oracleH. Our first aim will be to show that the advantage of any attacker-distinguisher pair, as defined above, is negligible at best. \\

\begin{figure}[!htb]
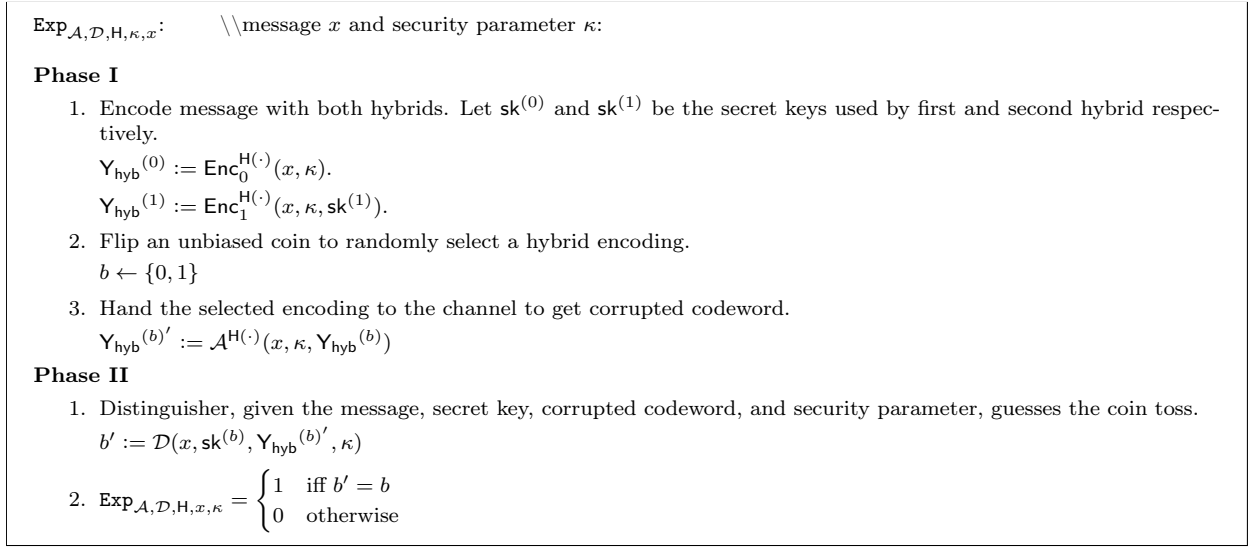

\begin{mdframed}
\footnotesize
$\mathtt{Exp}_{\attacker,\distinguisher,\mathsf{H},\kappa,x}$: \qquad\textbackslash\textbackslash message $x$ and security parameter \secpar:\\

\textbf{Phase I}
\begin{enumerate}
	\item Encode message with both hybrids. Let $\skHybridZero$ and $\skHybridOne$ be the secret keys used by first and second hybrid respectively.
	
	$\codewordHybridZero := \encHybridZero(x,\secpar)$. 
	
	$\codewordHybridOne := \encHybridOne(x, \secpar, \skHybridOne)$. 
	\item Flip an unbiased coin to randomly select a hybrid encoding.
	
	$b \leftarrow\{0,1\}$
	\item Hand the selected encoding to the channel to get corrupted codeword.
	
	$\corrcodewordHybridExp:= \roattacker(x,\secpar, \codewordHybridExp)$
\end{enumerate}
\textbf{Phase II}
\begin{enumerate}
    \item Distinguisher, given the message, secret key, corrupted codeword, and security parameter, guesses the coin toss.

    $b' := \distinguisher(x, \skHybridExp, \corrcodewordHybridExp, \secpar)$
	\item  $\mathtt{Exp}_{\attacker,\distinguisher,\mathsf{H},x,\secpar} =         \begin{cases}
                1 & \text{iff $b' = b$}\\
                0 & \text{otherwise}
        \end{cases}$
	
\end{enumerate}
\end{mdframed}
\vskip -0.1in
\caption{Indistinguishability experiment for the attacker-distinguisher pair.}
\label{fig:indist:exp}
\end{figure}

\myindent Let \ADpair be any attacker-distinguisher pair and hybrid encoders be instantiated with parameters \parameters{\safe, \codeLDC, \codePriv}. For security parameter \secpar and message $x$, consider an execution of the indistinguishability experiment \indistinguishExp. Let 
$\bad_\attacker$ be the event that the attacker queries the random oracle at position $c \circ \safe(\seedHybridExp)$ where $\controlinfo$ is the random seed chosen by the selected hybrid encoder \encHybridExp and $c$ is any constant expressed in binary. Furthermore, let \SUCCESS be the event where the attacker-distinguisher pair succeed in distinguishing the hybrid encodings in the experiment, i.e., the event where $\indistinguishExp = 1$\\

\myindent 
The next proposition follows from the observation that conditioning on the event $\bad_\attacker$ not occurring, the secret key $\sk_b$ used during the encoding process remains (information theoretically) private to both the adversary and the distinguisher.
To the pair, \encHybridZero appears information theoretically identical to \encHybridOne which gets a secret key as its input, and thus any advantage on distinguishing the encoding schemes would allow the pair to distinguish between random strings.

\begin{proposition} \label{obs:gamedistinguish_nbad}
$\Pr[\SUCCESS | \nbad_\attacker] = 1/2$
\end{proposition}
The following lemma shows that the advantage for any attacker-distinguisher pair is negligible. 
\begin{lemma} \label{lem:advAD_upper_bound}
$\advantageAD \leq \frac{\numOracleQueries\delta}{2}$ for any execution of the game $\mathtt{Exp}_{\attacker,\distinguisher,\mathsf{H},x,\secpar}$. Here $\numOracleQueries$ is an upper bound on the number of queries \roattacker makes to the random oracle.
\end{lemma}
\begin{proof}
Consider some execution of the game $\mathtt{Exp}_{\attacker,\distinguisher,\mathsf{H},x,\secpar}$. Using conditional probability to partition the event space, the advantage of the attacker-distinguisher pair is: $$\advantageAD = \bigg|\Pr[\SUCCESS] - \frac{1}{2} \bigg| = \bigg|\Pr[\SUCCESS | \bad_\attacker]\Pr[\bad_\attacker] + \Pr[\SUCCESS| \nbad_\attacker]\Pr[\nbad_\attacker] - \frac{1}{2} \bigg|$$

By Proposition~\ref{obs:gamedistinguish_nbad}, we may view the event of \SUCCESS conditioned on $\bad_\attacker$ not occurring as an unbiased random choice. 
Thus $\advantageAD =\bigg|\Pr[\SUCCESS | \bad_\attacker]\Pr[\bad_\attacker] + \frac 1 2 (1-\Pr[\bad_\attacker]) - \frac{1}{2} \bigg|$. 
This allows us to bound the advantage of the attacker-distinguisher pair by a factor of the probability of event \bad\ occurring by $\advantageAD=\Pr[\bad_\attacker] \bigg|\Pr[\SUCCESS | \bad_\attacker] - \frac{1}{2} \bigg|\leq\Pr[\bad_\attacker] \frac 1 2$. 
Therefore by Lemma~\ref{lem:prob_bad_event_attacker}, the advantage of the attacker-distinguisher pair for the execution of $\mathtt{Exp}_{\attacker,\distinguisher,\mathsf{H},x,\secpar}$ is at most $\frac{\numOracleQueries\delta}{2}$.

\ignore{
	\begin{align*}
		Adv(\distinguisher,\roattacker)& = \bigg|Pr[\SUCCESS] - \frac{1}{2} \bigg|&\\
		& =  \bigg|Pr[\SUCCESS | \bad_\attacker]Pr[\bad_\attacker] + Pr[b=b' | \nbad_\attacker]Pr[\nbad_\attacker] - \frac{1}{2} \bigg|&\\
		&=  \bigg|Pr[\SUCCESS | \bad_\attacker]Pr[\bad_\attacker] + \frac 1 2 (1-Pr[\bad_\attacker]) - \frac{1}{2} \bigg|& \ldots \text{ Observation } \ref{obs:gamedistinguish_nbad}\\
		&= Pr[\bad] \bigg|Pr[\SUCCESS | \bad_\attacker] - \frac{1}{2} \bigg|&\\
		&\leq Pr[\bad] \frac 1 2&\\
		&\leq \frac{q\eps}{2} & \ldots \text{ Lemma } \ref{lem:prob_bad_event_attacker}
	\end{align*}
	}
\end{proof}

\subsection{Security and Decoding Probability of Constructions}\label{sec:securityAndDecodingProb}

Note that \encHybridZero is identical to \encFinal and \encHybridOne is identical to \encPriv with random strings appended to its output.
Consider a $(\locPriv, \errorratePriv, \probrecovPriv, \epsilonPriv)-$private \LDC{} instance $\codePriv\parameters{\lenmsgPriv,\lencodewordPriv} = (\encPriv, \decPriv, \genkeyPriv)$ and an instantiation of our constructions $\codeFinal\parameters{\safe, \codeLDC,\codeFinal} = (\encFinal, \decFinal)$. With respect to these instances, we define $\epsilonFinal$ as the following:
$$\epsilonFinal := \Pr[\LDCsecgame = 1 \text{ against \codeFinal}] $$
Consider the codes $\codeHybridZero = (\encHybridZero, \decFinal)$ and $\codeHybridOne = (\encHybridOne, \decHybridOne)$ formed by our hybrid encoders. Here $\decHybridOne$ is defined identical to $\decPriv$ except that it ignores the strings appended to the output of $\encPriv$ during the encoding execution of \encHybridOne. With respect to these codes, we define the following:
$$\epsilonHybridZero := \max_{\roattacker \in \class} \Pr[\privLDCsecgameHybrid = 1 \text{ against \codeHybridZero}] $$
$$\epsilonHybridOne := \max_{\roattacker \in \class} \Pr[\privLDCsecgameHybrid = 1 \text{ against \codeHybridOne}] $$
Note that by our definitions, $\epsilonHybridZero = \epsilonFinal$ and $\epsilonHybridOne \leq  \epsilonPriv$. The second observation follows from the following:
\begin{align*}
    \epsilonHybridOne & = \max_{\roattacker \in \class}
    \Pr[\privLDCsecgameHybrid = 1 \text{ against } \codeHybridOne] &\\
     &\leq \max_{\roattacker \in \class} \Pr[\mathtt{{priv-LDC-Sec-Game}}\parameters{\attacker,\msg, \secpar, \errorrateFinal, \probrecovPriv} = 1 \text{ against } \codeHybridOne]  &\\
     & \leq \max_{\attacker \in \class} \Pr[\mathtt{{priv-LDC-Sec-Game}}\parameters{\attacker,\msg, \secpar, \errorratePriv, \probrecovPriv} = 1 \text{ against } \codePriv] = \epsilonPriv& 
\end{align*}
where the first inequality follows because  $\probrecovFinal \leq \probrecovPriv$, while the second inequality follows since $\errorrateFinal\lencodewordFinal \leq \errorratePriv\lencodewordPriv$ i.e., the attacker gets to make more corruptions against \codePriv. 
Lemma \ref{lem:distinguisher_contradiction} upper bounds $\left| \epsilonHybridZero - \epsilonHybridOne \right| \leq \numOracleQueries \delta$ and it immediately follows that $\epsilonFinal \leq \epsilonPriv + \numOracleQueries\delta$. 

\begin{lemma}\label{lem:distinguisher_contradiction}
$\left| \epsilonHybridZero - \epsilonHybridOne \right| \leq \numOracleQueries \delta$. Here $\numOracleQueries$ is an upper bound on the number of queries the attacker makes to the random oracle.
\end{lemma}
\begin{proof}
Recall that an attacker wins the \LDCsecgame if there exists some index which the corresponding decoder fails to decode with probability at least $p$. Suppose for sake of contradiction that $\left| \epsilonHybridZero - \epsilonHybridOne \right| > \numOracleQueries\delta$ for some attacker \roattacker. 
Consider the distinguisher $\mathsf{D'}$ in Figure~\ref{fig:distinguisher:ops}. With respect to the indistinguishability experiment, $\mathsf{D'}$ takes as input the original message \msg, the corrupted codeword $\corrcodeword_b$, the key used by hybrid $b$ during encoding, and the security parameter \secpar.

\begin{figure}[!htb]
\footnotesize
\begin{mdframed}
$\mathsf{Distinguisher \ D'}(\msg, \corrcodeword_b, \sk_b, \kappa)$:
\begin{enumerate}
    \item Computes $\epsilon_b$ by enumerating over all $i$, running $\decPriv^{\corrcodeword_b}(i, \secpar)$ and checking whether \decPriv fails to decode correctly with probability at least \probrecovPriv.
    \item return $b' = \begin{cases}
                    1 & \text{with probability $\epsilon_b$}\\
                    0 & \text{otherwise}
                    \end{cases}$
\end{enumerate}
\end{mdframed}
\vskip -0.1in
\caption{Distinguisher that uses the \decPriv decoding algorithm.}
\label{fig:distinguisher:ops}
\end{figure}
Note that the computationally intensive step $1$ of $\mathsf{D'}$ is possible since we assume no computational restrictions. 
Thus by conditional probability, the advantage of distinguisher $\mathsf{D'}$ paired with any $\roattacker \in \class$ may be given by
\begin{align*}
    \mathsf{Adv}_{\attacker,\mathcal{D}'}&= \abs{\Pr[\SUCCESS] - \frac 1 2}= \frac{1}{2}\abs{\Pr[\SUCCESS | b = 0] - \Pr[\nSUCCESS | b = 1]}\\
    &= \frac{1}{2}\abs{(1 - \epsilonHybridZero) - (1 - \epsilonHybridOne)}
    = \frac{1}{2}\abs{\epsilonHybridOne - \epsilonHybridZero},
\end{align*}
where the penultimate equality is by definition of the distinguisher $\mathsf{D'}$. 
Our initial assumption $\abs{\epsilonHybridZero - \epsilonHybridOne} > \numOracleQueries\delta$ then implies that $\advantageAD >  \frac{\numOracleQueries\delta}{2}$, contradicting Lemma \ref{lem:advAD_upper_bound}.
\end{proof}

The following proposition is a direct consequence of Lemma \ref{lem:distinguisher_contradiction} and the observation that $\epsilonHybridOne \leq \epsilonPriv$.

\begin{proposition} \label{prop:errorbound}
$\epsilonHybridZero \leq \epsilonPriv + \numOracleQueries\delta$ where $\numOracleQueries$ is an upper bound to the number of queries that the attacker makes to the random oracle.
\end{proposition}
Finally, we complete the proof by showing that  that $\epsilonFinal \leq \epsilonHybridZero$ in Lemma \ref{lem:epsilonfinal_prob_bound}.  Combined with  proposition \ref{prop:errorbound} this completes the proof since $\epsilonFinal \leq \epsilonHybridZero + \numOracleQueries \delta$.  
 
\begin{lemma} $\epsilonFinal \leq  \epsilonHybridZero$ \label{lem:epsilonfinal_prob_bound}
\end{lemma}
\begin{proof}
Let $\FAIL_i$ denote the event that \decFinal incorrectly decodes $x_i$ for $i \in [\lenmsg]$. We define $\SUCCESS$ to be the event that $\privLDCsecgameHybrid = 1 \text{ against \codeHybridZero}$ to simplify notation.   It suffices to argue that 
$\Pr[\FAIL_i | \nSUCCESS] \leq (1 - \probrecovPriv) + (1 - \probrecovLDC)$ for any $i \in [\lenmsg]$ since $\Pr[\SUCCESS] = \epsilonHybridZero$.
Let \KEY be the event that \decFinal recovers the correct seed $\seedHybridZero$ from $\ldccodewordHybridZero$.
We first observe that \begin{align*}
    \Pr[\FAIL_i | \nSUCCESS] &= \Pr[\FAIL_i | \nSUCCESS , \KEY]\Pr[\KEY | \nSUCCESS] + \Pr[\FAIL_i | \nSUCCESS , \nKEY]\Pr[\nKEY | \nSUCCESS]&\\
    &\leq \Pr[\FAIL_i | \nSUCCESS , \KEY] + \Pr[\nKEY | \nSUCCESS]&
\end{align*}

Second we observe that $\Pr[\nKEY | \nSUCCESS] \leq 1 - \probrecovLDC$  since there are at most $\errorrateFinal \lencodewordFinal \leq \errorrateLDC \lencodewordLDC$ errors in the second part of the codeword $\ldccodewordHybridZero$. Finally, observe that  by definition we have $\Pr[\FAIL_i | \nSUCCESS , \KEY]  \leq  1 - \probrecovPriv$. The claim now directly follows.
\end{proof}


\section{Constructing Safe Functions}
\label{sec:safe}
In this section we provide several examples of safe functions in the parallel random oracle model (\pROM) ~\cite{STOC:AlwSer15}. 
We first define the parallel random oracle model and introduce several cost metrics that measure the resources used by a \pROM algorithm \roattacker.

\subsection{Parallel Random Oracle Model} \label{sec:pROM}
Computation in the \pROM proceeds in rounds. 
Each round ends when the algorithm $\attacker$ outputs a batch of random oracle queries to be answered in parallel and a new round begins when the attacker receives the answer(s) to this batch of queries. 
In between rounds the $\attacker$ may perform arbitrary computation. 
Formally, in the initial round the \pROM algorithm $\attacker$ takes input $x$, performs some arbitrary computation, and outputs a state $\sigma_1$ and list $\vec{u}_1 = (u_1^1,\ldots, u_{\numOracleQueries_1}^1)$ of random oracle queries. 
In general, we then have $(\vec{u}_{i+1},\sigma_{i+1})= \attacker(\sigma_i, \vec{a}_i)$ where $\vec{a}_i= (\randOracleH{u_1^i},\ldots, \randOracleH{u_{\numOracleQueries_i}^i}$ are the answers to the $\numOracleQueries_i$ random oracle queries $\vec{u}_i = (u_1^i,\ldots, u_{\numOracleQueries_i}^i)$ asked in the previous round. 
The execution ends in round $t$ if the algorithm $\attacker$ returns an output value $y=\sigma_t$ along with an empty batch of random oracle queries $\vec{u}_t = \emptyset$. 
We use $$\etrace(x) = (\sigma_1, \sigma_2 \cdots , \sigma_t, \vec{u}_1,\ldots, \vec{u}_t )$$ to denote the sequence of states (and oracle queries) output when we run the \pROM attacker $\attacker(x)$ on input $x$ fixing the random oracle $\oracleH$ and fixing $\attacker$'s random coins $R$. 

\paragraph{Cost Metrics.}
Figure~\ref{tab:resource_notation} defines the \emph{resources} we will consider as characterizing the cost of a particular execution trace $\mathcal{T} = \etrace(x)$. 
We can define the time (resp. space) cost as $\restime(\mathcal{T})  =t$ (resp. $\resspace(\mathcal{T}) =\max_{i \leq t} |\sigma_i|$). 
Similarly, the space time cost measures the product $\resspacet(\mathcal{T})=t \cdot \max_{i \leq t} |\sigma_i|$ and cumulative memory complexity measures $\rescmc(\mathcal{T})=\sum_{i=0}^t |\sigma_i|$. 
Intuitively, cumulative memory complexity captures the amortized space time complexity of a function that we want to evaluate many times in parallel~\cite{STOC:AlwSer15}. 
Finally, the cumulative query cost is $\rescq(\mathcal{T}) = \sum_{i=1}^t |\vec{u_i}|$.

\begin{figure*}[htb]
\centering
\begin{tabular}{|c|c|c|}
 \hline
      & &  \\
   Resource  & Notation & Definition\\
   \hline
    & &\\
    Time  & $\restime(\mathcal{T})$ & $t$ \\
    & & \\
    Space  & $\resspace(\mathcal{T})$ & $\max_{i=0}^t|\sigma_i| $ \\
    &&\\
    Space-Time  & $\resst(\mathcal{T})$ & $\resspace(\mathcal{T})\cdot \restime(\mathcal{T})$\\
    &&\\
    Cumulative memory  & $\rescmc(\mathcal{T})$ & $\sum_{i = 0}^t |\sigma_i|$\\
    &&\\
    Cumulative query  & $\rescq(\mathcal{T})$ & $\sum_{i = 0}^{t} \vec{u}_i$\\
    &&\\
    \hline
\end{tabular}
\caption{Resource Definitions}
\label{tab:resource_notation}
\end{figure*}

\ignore{
\begin{center}
\begin{tabu} to 0.7\textwidth {| X | X | X |}
 \hline
      & &  \\
   Resource  & Notation & Definition\\
   \hline
    & &\\
    Time  & $\restime(\mathcal{T})$ & $t$ \\
    & & \\
    Space  & $\resspace(\mathcal{T})$ & $\max_{i=0}^t|\sigma_i| $ \\
    &&\\
    Space-Time  & $\resst(\mathcal{T})$ & $\resspace(\mathcal{T})\cdot \restime(\mathcal{T})$\\
    &&\\
    Cumulative memory  & $\rescmc(\mathcal{T})$ & $\sum_{i = 0}^t |\sigma_i|$\\
    &&\\
    Cumulative query  & $\rescq(\mathcal{T})$ & $\sum_{i = 0}^{t}p_i$\\
    &&\\
    \hline
\end{tabu}\\
\begin{tabu} to \textwidth { X  X  X }
&&\\
&Table 2.2\label{tab:resource_notation}: Resource Definitions&\\
\end{tabu}\\
\end{center}
}

\myindent
For a resource \res\ listed in Figure~\ref{tab:resource_notation}, the term \emph{\res\ complexity} will refer to a upper bound on resource \res. 

\begin{definition} (Resource Bounded Algorithms) We use $\mathcal{C}_{\rescq,\numOracleQueries}$ to refer to the set of all $\pROM$ algorithms $\mathcal{A}$ with the property that for all inputs $x$, random oracles $\oracleH{}$, and all random strings $R$, we have $\rescq(\etrace(x)) \leq \numOracleQueries$. We use $\mathcal{C}_{\resspace,M,} \subset \mathcal{C}_{\rescq,\numOracleQueries}$ to refer to the subset  of all $\pROM$ algorithms $\mathcal{A}$ with the additional constraint that for all inputs $x$, random oracles $\oracleH{}$, and all random strings $R$, we have $\rescq(\etrace(x)) \leq \numOracleQueries$ and $\resspace(\etrace(x)) \leq M$. Similarly, $\mathcal{C}_{\restime,T,\numOracleQueries} \subset \mathcal{C}_{\rescq,\numOracleQueries}$ (resp. $\mathcal{C}_{\resspacet,S,\numOracleQueries}\subset \mathcal{C}_{\rescq,\numOracleQueries}$) refers to the subset of all $\pROM$ algorithms $\mathcal{A}$ with the additional constraint that  for all inputs $x$, random oracles $\oracleH{}$, and all random strings $R$, we have  $\restime(\etrace(x)) \leq T$ (resp. $\resspacet(\etrace(x)) \leq S$). The definition of $\mathcal{C}_{\rescmc,M,\numOracleQueries}$ is symmetric --- we add the additional constraint that $\rescmc(\etrace(x)) \leq M$ for all $x,R,\oracleH{}$.
\end{definition}

The assumption that the channel is resource constrained with respect to one or more of the above resources (time, space, cmc,  etc.) is natural in most real word settings. 
For example, if a low latency channel uses $\roattacker$ to compute the corruptions to an encoded message then we can plausibly assume that the attacker $\attacker \in \mathcal{C}_{time,M,\numOracleQueries}$ is time bounded --- $M$ denotes the maximum number of sequential evaluations of $\oracleH$ before the corrupted codeword must be delivered. 
It would also be reasonable to assume that the total number of random oracle queries $\numOracleQueries$ is polynomial in the relevant parameters. 
One can also argue that in most practical settings the channel $\attacker$ will have other resource constraints e.g., space-bounded etc. 
In general one can define complexity classes for various combinations of resource constraints --- see Definition \ref{def:constraint_class}.

\begin{definition}\label{def:constraint_class}
For constraints $\constraints=(\constraints_1,\ldots,\constraints_p)$ on resources $\res{}=(\res_1,\ldots,\res_p)$ listed in Figure~\ref{tab:resource_notation}, the constraint class $C_{\res,\constraints}$ is the set of all \pROM \roattacker such that \roattacker is $\res$-bounded with constraints \constraints. Here, a \pROM algorithm is said to be \emph{$\res$-bounded} with \emph{constraints $\constraints$} if for all $i \leq p$ and on all inputs $x$, random coins $R$, and random oracles $\oracleH{}$, we have 
\[\res_i\big(\etrace (x)\big) \leq \constraints_i.\]
\end{definition}

\paragraph{SCRYPT.} 
Alwen et al.~\cite{EC:ACPRT17} proved that Percival's~\cite{Per09} memory hard function $\mathtt{scrypt}$ is maximally memory hard. 
In particular, $\mathtt{scrypt}_N$ can be computed in sequential time $N$, but any \pROM attacker evaluating the scrypt function has cumulative memory complexity at least $\Omega(N^2 w)$, where $w$ is the length of the output.  
Thus, $\mathtt{scrypt}$ could be used to obtain safe functions for the classes $\mathcal{C}_{\rescmc,S,\numOracleQueries}$ and  $\mathcal{C}_{\resspacet,S,\numOracleQueries}$ --- observe that $\rescmc(\mathcal{T}) \leq \resspacet(\mathcal{T})$ for any execution trace $\mathcal{T}$. 

\subsection{Sequentially Hard Function}
The hash iteration function $f(x)=\randOracleH{x}^{t+1}$, defined recursively as $\randOracleH{x}^{t+1} = \randOracleH{ \randOracleH{x}^t}$ where $\randOracleH{x}^1=\randOracleH{x}$, is a simple example of a safe function for the class $\mathcal{C}_{\restime,T=t,\numOracleQueries}$ of time bounded attackers --- see Claim \ref{claim:seqsafe}. 
The trade-off is sharp since it is trivial to compute $f(x)$ in sequential time $t+1$. 
This is a desirable property in our context since the encoder/decoder both need to compute $f(x)$ for a random input $x$.

\myindent
We remark that the proof of Claim \ref{claim:seqsafe} is very similar to an argument of Cohen and Pietrzak~\cite{EC:CohPie18}. 
Our bound is slightly tighter, but less general. 
Cohen and Pietrzak~\cite{EC:CohPie18} proved that any \pROM algorithm running in time $t$ can produce an arbitrary $H$-sequence with probability at most $\O{\frac{\numOracleQueries^2}{2^w}}$. 
We can reduce the bound to $\O{\frac{\numOracleQueries t}{2^w}}$ since the attacker needs to compute a specific $H$-sequence i.e., $L_1,\ldots,L_{t+1}$ with $L_i = \randOracleH{x}^i$. 
In general, we may have $\numOracleQueries \ll t$.

\begin{claim} \label{claim:seqsafe}
Let $f(x)=\randOracleH{x}^{t+1}$ and let $\epsilon = (t+1)t/2^{w+1} + (\numOracleQueries t+1) 2^{-w}$ then the function $f$ is $\epsilon$-safe for the class $\mathcal{C}_{\restime,T=t,\numOracleQueries}$.
\end{claim}
\begin{proof} (sketch)
Let $L_i:=\randOracleH{x}^i$. 
We remark that if $L_1,\ldots,L_{j-1}$ are all distinct then \[\PPr{L_j=\randOracleH{L_{j-1}} \in \{L_1,\ldots,L_{j-1}} \leq (j-1)2^{-w}.\]
Thus, the probability of the event $\mathtt{COL}$ that $L_i = L_j$ for some $1 \leq i < j \leq t+1$ is at most $2^{-w} \sum_{j=1}^{t+1} (j-1)= (t+1)t/2^{w+1}$. 
We say that a particular random oracle query $u$ in round $i$ is lucky if the output is $\randOracleH{u} = L_j$ but the label $L_{j-1}$ had not previously been observed as the output to any earlier random oracle query. 
If $i$ denotes the maximum index such that $L_i$ has been observed as a random oracle output, then the probability that a particular query $u$ is lucky is at most 
\[\Pr[\randOracleH{u} \in \{L_{i+2},\ldots, L_{t+1}\} | \overline{\mathtt{COL}}] = (t-i)2^{-w} \leq t2^{-w}.\] 
Conditioning on the event $\overline{\mathtt{COL}}$ that no collisions occur, we can apply union bounds to show that, except with probability $\numOracleQueries t/2^2$, there are no lucky queries. 
If there are no lucky queries, then after $t$ sequential rounds the output $L_{t+1} = f(x)$ can be viewed as uniformly random and the probability that the attacker outputs $f(x)$ is at most $2^{-w}$ in this case. 
\end{proof}

If we let $r$ denote the maximum number of sequential calls to $\oracleH{}$ that can be evaluated in a second\footnote{Bonneau and Schechter~\cite{USENIX:BonSch14} estimated that SHA256 can be evaluated $r \approx 10^7$ times per second on a single core processor} then we could set $t=r \times L_{max}$, where $L_{max}$ denotes the maximum latency of the channel. Note that the encoder/decoder would need require time marginally higher than the latency  $L_{max} + 1/r \approx L_{max}$ to compute $H^{t+1}(x)$.

\subsection{Graph Labeling Functions}
We first define a labeling function $f_{G,H}(x)$, given a graph $G$, a hash function $H$, and an input $x$. 
\begin{definition}
\label{def:labeling}
Given a DAG $G=(V=[N],E)$ and a random oracle function $H:\Sigma^*\rightarrow \Sigma^w$ over an alphabet $\Sigma$, we define the labeling of graph $G$ as $L_{G,H}:\Sigma^*\rightarrow\Sigma^*$.
In particular, given an input $x$ the $(H,x)$ labeling of $G$ is defined recursively by
\[L_{G,H,x}(v)=
\begin{cases}
H(v\circ x),&\indeg(v)=0\\
H\left(v\circ L_{G,H,x}(v_1)\circ\cdots\circ L_{G,H,x}(v_d)\right),&\indeg(v)>0,
\end{cases}\]
where $v_1,\ldots,v_d$ are the parents of $v$ in $G$, according to some predetermined lexicographical order. 
We define $f_{G,H}(x)= \{L_{G,H,x}(s) \}_{s \in \sinks(G)}$. 
If there is a single sink node $s_G$ then $f_{G,H}(x)=L_{G,H,x}(s_G)$. 
We omit the subscripts $G,H,x$ when the dependency on the graph $G$ and hash function $H$ is clear.
\end{definition}

The graph labeling function can be used to construct safe functions for several different classes of resource bounded adversaries. 
In particular, the resources necessary to compute $f_{G,H}$ in the \pROM are tightly linked to the black pebbling cost of the DAG $G$. 

\paragraph{Parallel Black Pebbling Game.} 
A legal (parallel) pebbling $P=(P_0,P_1,\ldots,P_t)$ of a DAG $G=(V,E)$ consists of a sequence of pebbling configurations $P_i \subseteq V$ --- representing the set of labels $L_{G,H,x}(v)$ which are stored in memory at time $i$. 
We start with no pebbles on the graph $P_0 = \emptyset$, and can remove pebbles from the graph (free memory) at any time. For any newly pebbled node $v \in P_{i+1}\setminus P_i$, it must be the case that $\parents(v) \subseteq P_i$ where $\parents(v):= \{ u ~:~(u,v) \in E\}$. Intuitively, this is because we cannot compute $L_{G,H,x}(v)$ unless each of the dependent values $L_{G,H,x}(u)$ for each $u \in \parents(v)$ is already available in memory. 
In the parallel version of the black pebbling game, there is no constraint on the number of new pebbles $\left| P_{i+1}\setminus P_i \right|$ that can be placed on the graph in each round.

\myindent
The space cost of a pebbling $P$ is defined as $\resspace(P):=\max_i |P_i|$ and the space complexity of a graph is $\resspace(G)=\min_{P} \resspace(P)$. 
The space-time (resp. cumulative cost) cost of a pebbling $P$ is the product $\resspacet(P):=\restime(P) \times \resspace(P)$ (resp. $\mathtt{CC}(G)=\sum_i |P_i|$). 
We remark that $\mathtt{CC}(G) \leq \resspacet(G)$.

For constant degree graphs $G$ with $N$ nodes it is known that $\resspace(G) = \O{N/\log N}$~ and that $\mathtt{CC}(G) = \O{N^2 \log \log N/\log N}$~\cite{C:AlwBlo16}. 
One can also construct graphs $G$ s.t. $\mathtt{CC}(G) = \Omega(N^2/\log N)$~\cite{EC:AlwBloPie17,CCS:AlwBloHar17} and Paul et al.~\cite{PTC76}  constructed a constant indegree graph $G$ with $\resspace(G) = \Omega(N/\log N)$~\cite{PTC76,EC:AlwBloPie18} --- this last bound is tight as Hopcroft et al.~\cite{HPV77} showed that {\em any} static DAG $G$ on $N$ nodes with constant indegree can be pebbled with at most $\resspace(G) = \O{N/\log N}$ pebbles. 

\paragraph{Pebbling Reductions.}  
In the appendix we prove that if $\resspace(G) \geq m$ and $S = mw/2$ then $f_{G,H}$ is safe for the class $\mathcal{C}_{\resspace, S,\numOracleQueries}$. The pebbling reduction is conceptually very similar to the reduction of Alwen and Serbinenko~\cite{STOC:AlwSer15} who proved that $\rescmc(f_{G,H}) = \Omega( \mathtt{CC}(G) \cdot w)$ i.e., if the graph $G$ has high cumulative pebbling cost then $f_{G,H}$ is safe for the class $\mathcal{C}_{\rescmc, M, \numOracleQueries}$, and by extension safe for the class $\mathcal{C}_{\resspacet, M, \numOracleQueries}\subseteq \mathcal{C}_{\rescmc, M, \numOracleQueries}$. In particular, given an execution trace $\etrace(x)$ for an algorithm $\attack(x)$ computing $f_{G,H}(x)$ we can (with high probability) extract a legal pebbling $P=(P_1,\ldots,P_t)$ for $G$ and then use an extractor argument to show that $\left| \sigma_i \right|/w \geq |P_i|/2$ during each round $i$ --- otherwise we could derive a contradiction by using the extractor to compress the random oracle. Thus, to construct a safe function one simply needs to find a graph $G$ with sufficiently large pebbling cost.

\subsection{Brief Note on Candidate Constructions without Random Oracles}
Recall that the proof of correctness for our \LDC{} constructions on space bounded channels uses the random oracle model inherently through an extractor argument showing that any space bounded channel that fools a decoding algorithm can also essentially predict a random string. 
However, we do not inherently require the random oracle model for general resource bounded channels. 
Thus in this section, we sketch candidate constructions for \LDCs{} on resource bounded channels that do not require the random oracle model.

\myindent
In the case where the channel must forward the (corrupted) codeword to the receiver within a certain amount of time, we can use other cryptographic primitives rather than a sequence of nested hash functions. 
For example, \emph{time-lock puzzles} \cite{rivest1996time} are designed so that a sender can quickly generate a puzzle with a solution that remains hidden until some predetermined amount of time has elapsed, even if an adversary has a polynomially large number of parallel processors. 
On the other hand, the solution is straightforward to calculate for any honest user who has spent the desired amount of time computing the puzzle. 
\cite{ITCS:BGJPVW16} propose time-lock puzzles through the use of succinct randomized encodings from indistinguishability obfuscation and the minimal assumption that ``inherently sequential'' languages exist. 

\myindent
For our purposes, an encoding algorithm can generate a time-lock puzzle whose solution is the random key and then transmit the time-lock puzzle along with the encoded message, using some repetition code to ensure that the time-lock puzzle can be determined by the decoding algorithm. 
The decoding algorithm can then solve the time-lock puzzle to obtain the random key and decode the message. 
However, if the channel is bounded by time $t$ and the hardness parameter of the time-lock puzzle is greater than $t$, then the channel cannot recover the random key. 
It is plausible that the same construction would also yield space-bound (or space-time bound) puzzles from minimal assumptions. 

\section*{Acknowledgements} We would like to thank anonymous reviewers for helpful feedback that improved the presentation of this paper. This research was supported in part by the National Science Foundation (CCF Award \#1910659).

\bibliographystyle{alpha}
\bibliography{abbrev0,crypto,references}

\newcommand{\etalchar}[1]{$^{#1}$}
\begin{thebibliography}{BFALS91}

\bibitem[AB16]{C:AlwBlo16}
Jo{\"e}l Alwen and Jeremiah Blocki.
\newblock Efficiently computing data-independent memory-hard functions.
\newblock In Matthew Robshaw and Jonathan Katz, editors, {\em Advances in
  Cryptology -- {CRYPTO}~2016, Part~II}, volume 9815 of {\em Lecture Notes in
  Computer Science}, pages 241--271, Santa Barbara, CA, USA, August~14--18,
  2016. Springer, Heidelberg, Germany.

\bibitem[ABH17]{CCS:AlwBloHar17}
Jo{\"e}l Alwen, Jeremiah Blocki, and Ben Harsha.
\newblock Practical graphs for optimal side-channel resistant memory-hard
  functions.
\newblock In Bhavani~M. Thuraisingham, David Evans, Tal Malkin, and Dongyan Xu,
  editors, {\em ACM CCS 2017: 24th Conference on Computer and Communications
  Security}, pages 1001--1017, Dallas, TX, USA, October~31~--~November~2, 2017.
  {ACM} Press.

\bibitem[ABP17]{EC:AlwBloPie17}
Jo{\"e}l Alwen, Jeremiah Blocki, and Krzysztof Pietrzak.
\newblock Depth-robust graphs and their cumulative memory complexity.
\newblock In Jean{-}S{\'{e}}bastien Coron and Jesper~Buus Nielsen, editors,
  {\em Advances in Cryptology -- {EUROCRYPT}~2017, Part~III}, volume 10212 of
  {\em Lecture Notes in Computer Science}, pages 3--32, Paris, France,
  April~30~--~May~4, 2017. Springer, Heidelberg, Germany.

\bibitem[ABP18]{EC:AlwBloPie18}
Jo{\"e}l Alwen, Jeremiah Blocki, and Krzysztof Pietrzak.
\newblock Sustained space complexity.
\newblock In Jesper~Buus Nielsen and Vincent Rijmen, editors, {\em Advances in
  Cryptology -- {EUROCRYPT}~2018, Part~II}, volume 10821 of {\em Lecture Notes
  in Computer Science}, pages 99--130, Tel Aviv, Israel, April~29~--~May~3,
  2018. Springer, Heidelberg, Germany.

\bibitem[ACP{\etalchar{+}}17]{EC:ACPRT17}
Jo{\"e}l Alwen, Binyi Chen, Krzysztof Pietrzak, Leonid Reyzin, and Stefano
  Tessaro.
\newblock Scrypt is maximally memory-hard.
\newblock In Jean{-}S{\'{e}}bastien Coron and Jesper~Buus Nielsen, editors,
  {\em Advances in Cryptology -- {EUROCRYPT}~2017, Part~III}, volume 10212 of
  {\em Lecture Notes in Computer Science}, pages 33--62, Paris, France,
  April~30~--~May~4, 2017. Springer, Heidelberg, Germany.

\bibitem[AS15]{STOC:AlwSer15}
Jo{\"e}l Alwen and Vladimir Serbinenko.
\newblock High parallel complexity graphs and memory-hard functions.
\newblock In Rocco~A. Servedio and Ronitt Rubinfeld, editors, {\em 47th Annual
  {ACM} Symposium on Theory of Computing}, pages 595--603, Portland, OR, USA,
  June~14--17, 2015. {ACM} Press.

\bibitem[BFALS91]{Babai_ldc}
László Babai, Lance Fortnow, Leonid A.~Levin, and Mario Szegedy.
\newblock Checking computations in polylogarithmic time.
\newblock pages 21--31, 01 1991.

\bibitem[BFNW91]{psuedo_random_gen_1}
L.~{Babai}, L.~{Fortnow}, N.~{Nisan}, and A.~{Wigderson}.
\newblock Bpp has subexponential time simulations unless exptime has
  publishable proofs.
\newblock In {\em [1991] Proceedings of the Sixth Annual Structure in
  Complexity Theory Conference}, pages 213--219, June 1991.

\bibitem[BGGZ19]{rldc_comp_bounded}
Jeremiah Blocki, Venkata Gandikota, Elena Grigorescu, and Samson Zhou.
\newblock Relaxed locally correctable codes in computationally bounded
  channels.
\newblock In {\em {IEEE} International Symposium on Information Theory,
  {ISIT}}, page (to appear), 2019.

\bibitem[BGH{\etalchar{+}}06]{Ben-SassonGHSV06}
Eli Ben{-}Sasson, Oded Goldreich, Prahladh Harsha, Madhu Sudan, and Salil~P.
  Vadhan.
\newblock Robust pcps of proximity, shorter pcps, and applications to coding.
\newblock {\em {SIAM} J. Comput.}, 36(4):889--974, 2006.
\newblock A preliminary version appeared in the Proceedings of the 36th Annual
  {ACM} Symposium on Theory of Computing (STOC).

\bibitem[BGJ{\etalchar{+}}16]{ITCS:BGJPVW16}
Nir Bitansky, Shafi Goldwasser, Abhishek Jain, Omer Paneth, Vinod
  Vaikuntanathan, and Brent Waters.
\newblock Time-lock puzzles from randomized encodings.
\newblock In Madhu Sudan, editor, {\em ITCS 2016: 7th Conference on Innovations
  in Theoretical Computer Science}, pages 345--356, Cambridge, MA, USA,
  January~14--16, 2016. Association for Computing Machinery.

\bibitem[BI01]{private_info_retrieval_1}
Amos Beimel and Yuval Ishai.
\newblock Information-theoretic private information retrieval: A unified
  construction.
\newblock In Fernando Orejas, Paul~G. Spirakis, and Jan van Leeuwen, editors,
  {\em Automata, Languages and Programming}, pages 912--926, Berlin,
  Heidelberg, 2001. Springer Berlin Heidelberg.

\bibitem[BS14]{USENIX:BonSch14}
Joseph Bonneau and Stuart~E. Schechter.
\newblock Towards reliable storage of 56-bit secrets in human memory.
\newblock In Kevin Fu and Jaeyeon Jung, editors, {\em USENIX Security 2014:
  23rd {USENIX} Security Symposium}, pages 607--623, San Diego, CA, USA,
  August~20--22, 2014. {USENIX} Association.

\bibitem[CKGS98]{private_info_retrieval_2_Chor}
Benny Chor, Eyal Kushilevitz, Oded Goldreich, and Madhu Sudan.
\newblock Private information retrieval.
\newblock {\em J. ACM}, 45(6):965--981, November 1998.

\bibitem[CMS99]{CachinMS99}
Christian Cachin, Silvio Micali, and Markus Stadler.
\newblock Computationally private information retrieval with polylogarithmic
  communication.
\newblock In {\em Advances in Cryptology - {EUROCRYPT} '99, International
  Conference on the Theory and Application of Cryptographic Techniques, Prague,
  Czech Republic, May 2-6, 1999, Proceeding}, pages 402--414, 1999.

\bibitem[CP18]{EC:CohPie18}
Bram Cohen and Krzysztof Pietrzak.
\newblock Simple proofs of sequential work.
\newblock In Jesper~Buus Nielsen and Vincent Rijmen, editors, {\em Advances in
  Cryptology -- {EUROCRYPT}~2018, Part~II}, volume 10821 of {\em Lecture Notes
  in Computer Science}, pages 451--467, Tel Aviv, Israel, April~29~--~May~3,
  2018. Springer, Heidelberg, Germany.

\bibitem[DGY11]{DvirGY11}
Zeev Dvir, Parikshit Gopalan, and Sergey Yekhanin.
\newblock Matching vector codes.
\newblock {\em {SIAM} J. Comput.}, 40(4):1154--1178, 2011.

\bibitem[DJK{\etalchar{+}}02]{self_correction1}
A.~{Deshpande}, R.~{Jain}, T.~{Kavitha}, S.~V. {Lokam}, and J.~{Radhakrishnan}.
\newblock Better lower bounds for locally decodable codes.
\newblock In {\em Proceedings 17th IEEE Annual Conference on Computational
  Complexity}, pages 184--193, May 2002.

\bibitem[DKW11]{DziembowskiKW11}
Stefan Dziembowski, Tomasz Kazana, and Daniel Wichs.
\newblock One-time computable self-erasing functions.
\newblock In {\em Theory of Cryptography - 8th Theory of Cryptography
  Conference, {TCC} Proceedings}, pages 125--143, 2011.

\bibitem[Efr12]{Efremenko12}
Klim Efremenko.
\newblock 3-query locally decodable codes of subexponential length.
\newblock {\em {SIAM} J. Comput.}, 41(6):1694--1703, 2012.

\bibitem[GLR{\etalchar{+}}91]{self_correction_2_Gemmel}
Peter Gemmell, Richard Lipton, Ronitt Rubinfeld, Madhu Sudan, and Avi
  Wigderson.
\newblock Self-testing/correcting for polynomials and for approximate
  functions.
\newblock In {\em Proceedings of the Twenty-third Annual ACM Symposium on
  Theory of Computing}, STOC '91, pages 33--42, New York, NY, USA, 1991. ACM.

\bibitem[GRR18]{GurRR18}
Tom Gur, Govind Ramnarayan, and Ron~D. Rothblum.
\newblock Relaxed locally correctable codes.
\newblock In {\em 9th Innovations in Theoretical Computer Science Conference,
  {ITCS}}, pages 27:1--27:11, 2018.

\bibitem[GS16]{Guruswami_Smith:2016}
Venkatesan Guruswami and Adam Smith.
\newblock Optimal rate code constructions for computationally simple channels.
\newblock {\em J. ACM}, 63(4):35:1--35:37, September 2016.

\bibitem[HO08]{HemenwayO08}
Brett Hemenway and Rafail Ostrovsky.
\newblock Public-key locally-decodable codes.
\newblock In {\em Advances in Cryptology - {CRYPTO} 2008, 28th Annual
  International Cryptology Conference, Proceedings}, pages 126--143, 2008.

\bibitem[HOSW11]{HemenwayOSW11}
Brett Hemenway, Rafail Ostrovsky, Martin~J. Strauss, and Mary Wootters.
\newblock Public key locally decodable codes with short keys.
\newblock In {\em 14th International Workshop, {APPROX}, and 15th International
  Workshop, {RANDOM}, Proceedings}, pages 605--615, 2011.

\bibitem[HPV77]{HPV77}
John Hopcroft, Wolfgang Paul, and Leslie Valiant.
\newblock On time versus space.
\newblock {\em J. ACM}, 24(2):332--337, April 1977.

\bibitem[{Jus}72]{Justesen}
J.~{Justesen}.
\newblock Class of constructive asymptotically good algebraic codes.
\newblock {\em IEEE Transactions on Information Theory}, 18(5):652--656, Sep.
  1972.

\bibitem[KdW04]{KerenidisW04}
Iordanis Kerenidis and Ronald de~Wolf.
\newblock Exponential lower bound for 2-query locally decodable codes via a
  quantum argument.
\newblock {\em J. Comput. Syst. Sci.}, 69(3):395--420, 2004.

\bibitem[KMRS17]{KoppartyMRS17}
Swastik Kopparty, Or~Meir, Noga Ron{-}Zewi, and Shubhangi Saraf.
\newblock High-rate locally correctable and locally testable codes with
  sub-polynomial query complexity.
\newblock {\em J. {ACM}}, 64(2):11:1--11:42, 2017.

\bibitem[KO97]{private_info_retrieval_3}
E.~{Kushilevitz} and R.~{Ostrovsky}.
\newblock Replication is not needed: single database, computationally-private
  information retrieval.
\newblock In {\em Proceedings 38th Annual Symposium on Foundations of Computer
  Science}, pages 364--373, Oct 1997.

\bibitem[KT00]{fault_tolerant_storage}
Jonathan Katz and Luca Trevisan.
\newblock On the efficiency of local decoding procedures for error-correcting
  codes.
\newblock In {\em Proceedings of the Thirty-second Annual ACM Symposium on
  Theory of Computing}, STOC '00, pages 80--86, New York, NY, USA, 2000. ACM.

\bibitem[Lip94]{lipton_new_1994}
Richard~J. Lipton.
\newblock A new approach to information theory.
\newblock In {\em {STACS} 94}, pages 699--708, Berlin, Heidelberg, 1994.

\bibitem[MPSW05]{micali_etal_computationally_bounded_noise}
Silvio Micali, Chris Peikert, Madhu Sudan, and David~A. Wilson.
\newblock Optimal error correction against computationally bounded noise.
\newblock In Joe Kilian, editor, {\em Theory of Cryptography}, pages 1--16,
  Berlin, Heidelberg, 2005. Springer Berlin Heidelberg.

\bibitem[OPS07]{OPS}
Rafail Ostrovsky, Omkant Pandey, and Amit Sahai.
\newblock Private locally decodable codes.
\newblock In {\em Automata, Languages and Programming}, pages 387--398, 2007.

\bibitem[Per09]{Per09}
C.~Percival.
\newblock Stronger key derivation via sequential memory-hard functions.
\newblock In {\em BSDCan 2009}, 2009.

\bibitem[PTC76]{PTC76}
Wolfgang~J. Paul, Robert~Endre Tarjan, and James~R. Celoni.
\newblock Space bounds for a game on graphs.
\newblock In {\em Proceedings of the Eighth Annual ACM Symposium on Theory of
  Computing}, STOC '76, pages 149--160, New York, NY, USA, 1976. ACM.

\bibitem[RSW96]{rivest1996time}
Ronald~L Rivest, Adi Shamir, and David~A Wagner.
\newblock Time-lock puzzles and timed-release crypto.
\newblock 1996.

\bibitem[SS16]{ShaltielS16}
Ronen Shaltiel and Jad Silbak.
\newblock Explicit list-decodable codes with optimal rate for computationally
  bounded channels.
\newblock In {\em Approximation, Randomization, and Combinatorial Optimization.
  Algorithms and Techniques, {APPROX/RANDOM}}, pages 45:1--45:38, 2016.

\bibitem[STV01]{psuedo_random_gen_2}
Madhu Sudan, Luca Trevisan, and Salil Vadhan.
\newblock Pseudorandom generators without the xor lemma.
\newblock {\em Journal of Computer and System Sciences}, 62(2):236 -- 266,
  2001.

\bibitem[Yek08]{Yekhanin08}
Sergey Yekhanin.
\newblock Towards 3-query locally decodable codes of subexponential length.
\newblock {\em J. {ACM}}, 55(1):1:1--1:16, 2008.

\end{thebibliography}

\appendix

\section{Repetition with Justesen Codes}\label{appendix:justesen}
As a preliminary to this section, we require familiarity with the following form of standard one-sided Chernoff Bounds.
\begin{proposition}[Chernoff Bound]\label{prop:Chernoff}
Let $\X_1, \cdots \X_n$ be \emph{independent} random variables such that $0 \leq \X_i \leq 1$ for each $i \in [n]$. Let $\S_n = \sum_{i = 1}^n \X_i$ and $\mu = \E[\S_n]$. Then for any $\epsilon > 0$,
$$\Pr\left[\S_n \leq (1 - \epsilon)\mu\right] \leq \exp\bigg(-\frac{\epsilon^2}{2}\mu\bigg)$$
\end{proposition}

In this section, we will describe our encoding scheme in order to recover the random seed \controlinfo used by our main constructions in Section~\ref{sec:framework}. Recall that our approach was to encode \controlinfo using a repetition code by repeating \controlinfo multiple (\numjrepcontrolcode) times, and then encoding each repetition of \controlinfo into \jcontrolcode\ using an off-the-shelf error correcting code with constant error and information rates.
While any constant rate error correcting code may be used, we make use of \emph{Justesen Codes}:

\begin{theorem}
{\cite{Justesen}}\label{thm:justesen_codes}
For any $0 < R_\mathsf{J} < 1$, there exist binary linear codes of rate $R_\mathsf{J}$, that are efficiently decodable from $\delta_\mathsf{J}(R_\mathsf{J})$ fraction of errors, where $\delta_\mathsf{J}$ is a function that only depends on $R_\mathsf{J}$.
\end{theorem}

We will denote $\mathsf{C_J} = (\justesenenc,\justesendec)$ as the code that achieving the guarantees of Theorem \ref{thm:justesen_codes}, i.e. having constant rate $R_\mathsf{J}$ and error correction rate $\delta_\mathsf{J}(R_\mathsf{J})$. 
Let \lenjcontrolcode\ denote the length of these codewords. 
We now give our code constructions to recover \controlinfo via repetition with Justesen codes.
\begin{mdframed}
\textbf{\Encjrep}(\controlinfo):
\begin{enumerate}
    \item $\jcontrolcode := \justesenenc(\controlinfo)$.
    \item $\jrepcontrolcode := \jcontrolcode \circ \jcontrolcode \circ \ldots \circ \jcontrolcode$ where \jcontrolcode\  is repeated \numjrepcontrolcode times for a fixed \numjrepcontrolcode.
    \item Output \jrepcontrolcode
\end{enumerate}
\end{mdframed}

\begin{mdframed}
\textbf{\Decjrep}($\jrepcontrolcode'$):
\begin{enumerate}
    \item $I :=$ Sample $\alpha$ indices uniform with replacement from $[\numjrepcontrolcode]$ for some prespecified $\alpha$.
    \item For $i \in I$ 
    
     \hspace*{4mm}$\jcontrolcode'^{(i)} := \jrepcontrolcode'[i\lenjcontrolcode, \ldots , (i+1)\lenjcontrolcode - 1]$ 
     
     \hspace*{4mm}$\controlinfo'^{(i)} := \justesendec(\jcontrolcode'^{(i)})$.
    \item Output $\majority(\controlinfo'^{(1)}, \controlinfo'^{(2)}, \cdots, \controlinfo'^{(\alpha)})$
\end{enumerate}
\end{mdframed}

The following lemma states that the code $(\Encjrep,\Decjrep)$ may be used to recover the original \controlinfo with high probability using good locality.
\begin{lemma}\label{lem:jrep} Let $\mathsf{C_J}\parameters{\mathsf{R_J}}$ be as in Theorem \ref{thm:justesen_codes} and $\alpha$ be the number of samples that \Decjrep makes to the corrupted codeword. Then
for all $k_{\mathsf{JREP}} > 0$, the $(K_{\mathsf{JREP}}, k_{\mathsf{JREP}})_2$ coding scheme $\mathsf{C_{JREP}}\parameters{k_{\mathsf{JREP}}, \mathsf{C_{J}}, \alpha} = (\Encjrep,\Decjrep)$ is an $(\jrepquery, \rhojrep, p_{\mathsf{JREP}})$-\LDCvariant{} where $\jrepquery = \alpha \frac{k_{\mathsf{JREP}}}{\mathsf{R_J}}$, $\rhojrep$ is some constant, and $p_{\mathsf{JREP}} \geq 1 - e^{-\alpha / 24}$.
\end{lemma}
\begin{proof}
The adversary makes a total of $\rhojrep\cdot\lenjrep$ corruptions to \jrepcontrolcode. 
Let $\rhoj = \delta_\mathsf{J}(R_\mathsf{J})$. 
By Theorem \ref{thm:justesen_codes}, for a \jcontrolcode\ block to be non-decodable by \justesendec, the adversary must make at least $\rhoj\lenjcontrolcode$ corruptions in this block. 
This allows us to bound the total blocks the adversary may corrupt as at most $\frac{\rhojrep\lenjrep}{\rhoj\lenjcontrolcode} = \frac{\rhojrep}{\rhoj}\numjrepcontrolcode $. 
Thus the probability of sampling a block that is non-decodable is at most $\frac{\rhojrep}{\rhoj}$. Setting $\rhojrep = \frac{\rhoj}{4}$, we get that the probability of sampling a block that may be recovered is at least $3/4$. 
Let $\S_\alpha$ denote the number of samples that are successfully recovered. Thus, $\E[\S_n] \geq \frac{3\alpha}{4}$ and by standard Chernoff Bounds (Proposition \ref{prop:Chernoff}), we have that 
$$\Pr\left[\S_\alpha \leq \frac \alpha 2\right] \leq \exp\big(-\frac{\alpha}{24}\big)$$

Note that if $\S_\alpha > \alpha/2$, then our majority vote succeeds in determining the original message. Thus $p_{\mathsf{JREP}} = \Pr[\S_\alpha > 1/2]$.
Finally, each block has size $\frac{k_{\mathsf{JREP}}}{\mathsf{R}_{\mathsf{J}}}$. 
Since we sample $\alpha$ blocks, we get the claimed locality.
\end{proof}

\section{Memory Bounded Adversary}
In this section we show that the memory complexity of the function $f_{G,H}$ is characterized by the space cost $\spacepeb(G)$ in the parallel random oracle model just as Alwen and Serbinenko \cite{STOC:AlwSer15} showed that cumulative memory complexity can be characterized by the black pebbling game. 

\subsection*{Graph Pebbling}
Given a directed acyclic graph (DAG) $G=(V,E)$, the goal of the (parallel) black pebbling game is to place pebbles on all sink nodes of $G$ (not necessarily simultaneously). 
The game is played in rounds and we use $P_i \subseteq V$ to denote the set of currently pebbled nodes on round $i$. 
Initially all nodes are unpebbled, $P_0 = \emptyset$, and in each round $i \ge 1$ we may only include $v \in P_i$ if all of $v$'s parents were pebbled in the previous configuration ($\parents(v) \subseteq P_{i-1}$) or if $v$ was already pebbled in the last round ($v \in P_{i-1}$). 

\myindent 
The cumulative cost of the pebbling is defined to be $|P_1|+\ldots + |P_t|$. 
Graph pebbling is a particularly useful as a tool to analyze the security of an iMHF~\cite{STOC:AlwSer15}. 
A pebbling of $G$ naturally corresponds to an algorithm to compute the iMHF. Alwen and Serbinenko~\cite{STOC:AlwSer15} proved that in the parallel random oracle model (\pROM) of computation, any algorithm evaluating such an iMHF could be reduced to a pebbling strategy with (approximately) the same cumulative memory cost. 

\myindent 
However for our purposes, we are more concerned about the space cost rather than the cumulative memory cost. 
The space of the pebbling is defined to be $\space(P)=\max_i|P_i|$ and accordingly, $\space(G)=\min\space(P)$, where the minimum is taken over all valid pebblings $P$. 

\subsection*{Reduction}
Similar to \cite{STOC:AlwSer15} our reduction uses Lemma~\ref{lem:hintsize} as a core building block. 
In particular, if the space complexity is significantly smaller than $\spacepeb(G)$ for a \pROM attacker then we will be able to build an extractor that receives a small hint and predicts the random oracle output on an index contradicting Lemma~\ref{lem:hintsize}.  
By contrast, a black pebbling move always corresponds to a specific random oracle query. 

\begin{lemma}\cite{DziembowskiKW11}
\label{lem:hintsize}
Let $B$ be a series of random bits and let $\mathcal{A}$ be an algorithm that receives a hint $h\in H$ and can query $B$ at specific indices. 
If $\mathcal{A}$ outputs a subset of $k$ indices of $B$ that were previously not queried, as well as guesses for each of the bits, the probability there exists some $h\in H$ so that all the $k$ guesses are correct is at most $\frac{|H|}{2^k}$.
\end{lemma}

\subsection{Memory and Cache in the Parallel Random Oracle Model}
Before we present our reduction, we first recall the formal definition of space complexity in the \pROM model. 
Let the state of an algorithm $\attack$ at time $i$ to be $\sigma_i$, which contains the contents of the memory. 
Let $\attack$ be a \pROM attacker $\attack$ who is given oracle access to a random oracle $H:\{0,1\}^{*}\rightarrow \{0,1\}^w$. 
An execution of  $\attack$ on input $x$ proceeds in rounds as follows. 
Initially, the state at time $0$ is $\sigma_0$, which encodes the initial input $x$. 
At the beginning of round $i$ the attacker is given the initial state $\sigma_{i-1}$ as well as the answers $A_{i-1}$ to any random oracle queries that were asked at the end of the last round. 
The algorithm $\attack$ may then perform arbitrary computation and choose to update the memory, outputting a new state $\sigma_i$, along with a batch of queries $Q_i=\{q^i_1,q^i_2,\ldots,q_i^{k_i}\}$. 

\paragraph{Execution Trace.}
Recall that the execution trace of the algorithm $\attack$ is defined by the sequence of memory states and queries made to the random oracle $H$. 
Formally, the execution trace is $\etrace(\attacker,x)=\{(\sigma_i,Q_i)\}_{i=1}^t$, where the trace $\etrace(\attacker,x)$ is dependent on the algorithm $\attack$, random oracle $H$, internal randomness $R$, and input value $x$. 
Then the memory cost of the execution trace is
\[\mcost(\etrace(\attacker,x))=\max_i|\sigma_i|\ .\]

\myindent 
Recall that Alwen and Serbinenko~\cite{STOC:AlwSer15} show that the computation of a function $f_{G,H}$ with hash function $H$ and underlying directed acyclic graph $G$ yields a legal black pebbling with high probability. 
Thus, we use $\attack$ to extract a legal $P = \left(P_1,\ldots,P_t\right) \in \pPeb(G)$.
Given an execution trace $\etrace(x)$, the corresponding pebbling $\mathtt{BlackPebble}^H\left(\etrace(x)\right)=P_0,\ldots,P_t$ is defined by setting $P_0=\emptyset$ and define the pebbles at each subsequent time step $i$ based on the corresponding batch of queries $q_i$ made during iteration $i$. 
We then apply the following rules:
\begin{itemize}
\item
For each query $q$ in batch $q_i$, if the query has the form $v,\lab_{H,x}(v_1),\ldots,\lab_{H,x}(v_d)$ for some vertex $v$ and its parents $v_1,\ldots,v_d$, then we add a pebble to node $v$ in $P_i$. 
\item
If there exists another query for $v$ before $v$ is used as input for a query, then $v$ is deleted from $P_i$. 
\end{itemize}
Intuitively, at each time $j$, $P_j$ contains all nodes $v$ whose label will appear as input to a future random oracle query {\em before} the label appears as the output of a random oracle query. 
In this manner, we define $\mathtt{BlackPebble}^H \left( \etrace(x)\right) = P_1,\ldots,P_t \subseteq V$, which Alwen and Serbineneko show is legal with high probability:
\begin{theorem}\cite{STOC:AlwSer15}
\label{thm:informal:black}
The pebbling extracted from an execution trace, 
\[\mathtt{BlackPebble}^H \left(\etrace(x)\right)\in \pPeb(G),\]
is a legal black pebbling with probability at least $1-\frac{q}{2^w}$, where $w$ is the label size and $q$ is the number of queries made by $\etrace$.
\end{theorem}

\myindent 
We now show that any algorithm $\attack$ that computes $f_{G,H}(x)$ correctly with probability at least $\eps$ has memory cost $\mcost$ dependent on the space complexity of the resulting legal black pebbling, $\spacepeb(\pPeb(G))$. 
The proof uses that fact that if an attacking strategy does not yield a corresponding legal black pebbling, then the attacking strategy can be modified to form an extractor for the labels of a subset of nodes. 
Specifically, an extractor with access to the attacking strategy, the state of the memory, and a few select hints can successfully predict a large number of random bits, which cannot happen with high probability. 
The hints given to the extractor describes the positions of the random bits, and ensure these bits remain ``random'' (that is, we do not explicitly query these locations later). 
In particular, the extractor uses the hints to simulate $\attack$ but the hints do not include the current state of memory $\sigma_i$. 




\begin{theorem}
\label{thm:reduction}
Let $G$ be a DAG with $n$ nodes, $w>8\log n$, $q<2^{w/16}$, and $x$ be a fixed input. 
Let $m=\mcost(\etrace(\attacker,x))$. 
For any algorithm $\mathcal{A}^{H(\cdot)}$ that makes at most $q$ queries, let $\HIGH(\attacker,x)$ be the event that the attacker either uses $\frac{w}{2}\spacepeb(G)$ in its computation of $f_{G,H}(x)$ or fails to compute the function correctly. 
Then
\[\PPr{\HIGH(\attacker,x)}\ge 1-\frac{q}{2^w}-\frac{1}{2^{-3mw/4}}-\frac{n^2}{2^{w+1}}.\]
\end{theorem}
\begin{proof}
Consider an instance of $\mathcal{A}^{H(\cdot)}$ that succeeds in calculating $f_{G,H}$, making at most $q$ queries. 
By Theorem~\ref{thm:informal:black}, with probability at least $1-\frac{q}{2^w}$, we can extract a legal black pebbling from $\mathcal{A}^{H(\cdot)}$. 
Conditioned on the success of the extraction of a legal black pebbling, let $m=\mcost(\etrace(\attacker,x))$ be the space complexity of the execution trace of the evaluation algorithm. 
By definition of $\spacepeb$, there exists a time step $i$ such that the corresponding legal black pebbling contains at least $m$ pebbles. 
Suppose, by way of contradiction, that $\mcost_{q,\eps}(f_{G,H})<\frac{mw}{2}$ so that $|\sigma_i|<\frac{mw}{2}$. 
By construction of the pebbling, there is a set $S$ containing $m$ labels that appear as input for a query after time step $i$ before they are returned as output. 
Moreover, there are collisions among labels with probability at most $\sum_{i=1}^n \frac{(i-1)}{2^w} \leq  \frac{n^2}{2^{w+1}}$, so conditioned on the event that there are no collisions among the labels, then the algorithm $\mathcal{A}^{H(\cdot)}$ would have to generate the labels out of thin air. 
Specifically, an extractor using $\mathcal{A}^{H(\cdot)}$ will be able to predict $m$ labels, each of size $w$ bits, using a $\frac{mw}{2}$ bits of information from the state of $\mathcal{A}^{H(\cdot)}$, along with the following hint, which consists of three parts:
\begin{enumerate}
\item
The set $S$ is given as a hint to denote the indices that form the string that the extractor will ultimately predict. 
Since $S$ contains $m$ positions, then the size of this component of the hint is $m\log n$ bits.
\item
For each $v\in S$, the index of the first query that appears in which $\lab(v)$ is needed as input.  
This component of the hint tells the extractor the queries that require the prediction of random strings, and has size at most $m\log q$ bits, where $q$ is the total number of queries made by the attacker.
\item
For each $v\in S$, the index of the first query when $\lab(v)$ might be compromised. 
Observe that if the extractor successfully predicts a random string at a location $v$, but then $\lab(v)$ is later queried by the attacker, we cannot distinguish this case at the end from the case that the extractor simply read $\lab(v)$ after making the query. 
Effectively, the extractor is no longer predicting a random string. 
To avoid this, the hint given to the extractor details queries that would compromise the randomness of the desired locations. 
Formally, the hint is the minimal index $i$ such that $q_i^j=v$, which yields returns the query $H(q_i^j)=\lab(v)$. 
This component of the hint tells the extractor the locations of the random strings to be predicted, and has size at most $m\log q$ bits.
\end{enumerate}
See Figure~\ref{fig:extractor} for intuition.
\begin{figure*}[htb]
\centering
\begin{tikzpicture}[scale=0.4]

\draw (-10cm,0cm) rectangle+(5cm,4cm);
\node at (-8cm,1cm){\tiny{Attacker $\mathcal{A}$}};
\draw  (-9.5cm,2cm) rectangle+(4cm,1.5cm);
\node at (-7.5cm,2.4cm){\tiny{Memory: $\sigma_i$}};
\node at (-3.5cm,4cm){$H(\cdot)$};

\node at (-1.5cm,4cm){$\longrightarrow$};
\draw (0.7cm,0cm) rectangle+(5cm,4cm);
\node at (3cm,1cm){\tiny{Attacker $\mathcal{A}$}};
\draw  (1cm,2cm) rectangle+(4cm,1.5cm);
\node at (3cm,2.4cm){\tiny{Memory: $\sigma_i$}};
\draw (0cm,-1cm) rectangle+(8.5cm,8cm);
\draw  (6.2cm,0.25cm) rectangle+(2cm,5cm);
\node at  (7cm,2.5cm){\rotatebox{90}{\tiny{Hint: $\sigma_i, \ldots$}}};
\draw  (0.5cm,4.5cm) rectangle+(3.5cm,2cm);
\node at (2cm,6cm){\tiny{RO Pairs:}};
\node at (2cm,5.5cm){\tiny{$(x,H(x))$}};
\node at (6cm,6cm){\tiny{Extractor}};
\node at (10cm,4cm){$H(\cdot)$};
\end{tikzpicture}
\caption{An extractor that uses the attacker to predict $m$ distinct outputs of random oracle $H(\cdot)$.}
\label{fig:extractor}
\end{figure*}
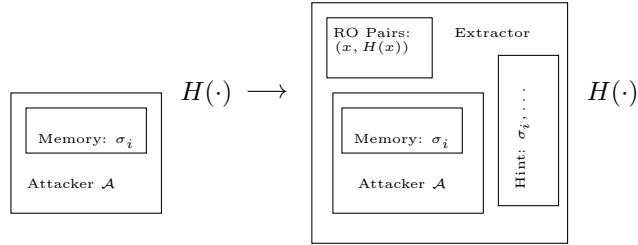
The size of the hint is at most $m\log n+2m\log q$ bits. 
However, the extractor is able to use $\mathcal{A}^{H(\cdot)}$ to predict $m$ fresh input/output pairs $(x_i,(H(x_i))$ from the random oracle. 
That is, the extractor can predict $m$ labels of length $w$ from the random oracle for a total of $mw$ random bits from $\frac{mw}{2}+m\log n+2m\log q$ bits.
Thus for $\log n<\frac{w}{8}$ and $q<2^{w/16}$, the extractor predicts $mw$ random bits from $\frac{3}{4}mw$ random bits, which can only occur with probability $\frac{1}{2^{-3mw/4}}$ by Lemma~\ref{lem:hintsize}. 
Hence, the probability that the attacker either uses $\frac{w}{2}\spacepeb(G)$ in its computation of $f_{G,H}(x)$ or fails to compute the function correctly is at least $1-\frac{q}{2^w}-\frac{1}{2^{-3mw/4}}-\frac{n^2}{2^{w+1}}$, where the possible events of failure are the inability to extract a legal pebbling from the attacker, the probability of extracting $mw$ random bits from $\frac{3}{4}mw$ random bits, and collisions among the labels, respectively. 
\end{proof}

\section{Code Scrambling}\label{appendix:codescrambling}

\newcommand{\mask}{\ensuremath{\mathsf{m}}\xspace}
\newcommand{\perm}{\ensuremath{\mathsf{\pi}}\xspace}
\newcommand{\symenc}{\ensuremath{\mathsf{Enc}_{\mathsf{SC}}}\xspace}
\newcommand{\symdec}{\ensuremath{\mathsf{Dec}_{\mathsf{SC}}}\xspace}
\newcommand{\symcode}{\ensuremath{\mathsf{C}_{\mathsf{SC}}}\xspace}
\newcommand{\symcodeword}{\ensuremath{y_{\mathsf{SC}}}\xspace}

\newcommand{\advenc}{\ensuremath{\mathsf{Enc}_{\mathsf{Adv}}}\xspace}
\newcommand{\advdec}{\ensuremath{\mathsf{Dec}_{\mathsf{Adv}}}\xspace}
\newcommand{\advcode}{\ensuremath{\mathsf{C}_{\mathsf{Adv}}}\xspace}

Code scrambling was a technique introduced by Lipton \cite{lipton_new_1994} for transforming codes designed for the symmetric channel, to be used against any \PPT\  adversarial channel. Assume that the sender and receiver share some private randomness $(\perm,\mask)$. Here \perm is a random permutation on $\{1, \cdots,
n\}$ and $\mask \in \{0,1\}^n$ is a random mask. $n$ is then length of the encoding obtained by  $\symcode = (\symenc, \symdec)$, a constant rate code in the symmetric channel. Consider the following code against the \PPT\   adversarial channel: given message $x$,

\begin{mdframed}
\advenc(x, \perm, \mask):
\begin{enumerate}
    \item $y_{\mathsf{SC}} := \symenc(x)$
    \item $y := \perm(y_{\mathsf{SC}}) \oplus \mask$
    \item Output $y$
\end{enumerate}
\end{mdframed}

\begin{mdframed}
\advdec(y', \perm, \mask):
\begin{enumerate}
    \item $y'_{\mathsf{SC}} := \perm^{-1}(y' \oplus \mask)$
    \item $x' := \symdec(y'_{\mathsf{SC}})$
    \item Output $x'$
\end{enumerate}
\end{mdframed}

The key observation is that \perm and $y_{\mathsf{SC}}$ are independent due to \mask. This may be observed by considering a specific \perm and realizing that the final encoding may take on any values due to the \mask. Thus if the error vector added by the adversarial channel is $\mathcal{E}$, then we have 
$$\symcodeword' = \pi^{-1}(y + \mathcal{E} + \mask) = \perm^{-1}(y + \mask) + \perm^{-1}(\mathcal{E}) = \symenc(x) + \perm(\mathcal{E})$$
Thus the errors added are random, due to the random permutation, and \symdec may recover the original message. 

\section{Private \LDCs}\label{appendix:privateldc}

\emph{Private locally decodable} codes were introduced by Ostrovsky, Pandey and Sahai \cite{OPS}. These {\LDC}s are termed \emph{private} as they crucially assume a \emph{secret key} given to both the sender and the receiver before the protocol, but kept private from the \PPT\ channel.

\begin{definition}\label{def:private-LDCs}
Let \secpar be the security parameter. A \emph{private $\ell$-locally decodable code} for a parameters $(K,k)$, is a triplet of probabilistic polynomial time algorithms $(\mathsf{GenKey}, \Enc, \Dec)$ such that:
\begin{itemize}
    \item $\mathsf{GenKey}(\secpar)$ is the key generation algorithm that takes as input the security parameter \secpar and outputs a secret key \sk.
    \item $\Enc(x,\sk)$ is the encoding algorithm that takes as input the message $x$ of length $k = \poly(\secpar)$ and the secret key \sk. The algorithm outputs $y \in \{0, 1\}^K$ that denotes an encoding of $x$.
    \item $\Dec(j,\sk)$ denotes the decoding algorithm, which takes as input a bit position $j \in [k]$ and the secret key \sk. It outputs a single bit $b$ denoting the decoding of $x[j]$ by making at most $\ell$ (adaptive) queries into a given a codeword $y'$ possibly different from $y$.
\end{itemize}
\end{definition}

 Here, $\ell$ denotes the query complexity or locality of the code and $\rho$ is termed the error rate. 
Furthermore, we say that the private \LDC\ decodes with probability $p$ if for \PPT\ channels in the experiment of definition \ref{def:private_ldc_game} (defined below) for all $x \in \{0,1\}^k$ and $i \in [k]$, we have $\Pr[b = x^{(h)}_i] \geq p$\\

The game between the encoder/decoder and \PPT\ adversarial channel may be described as follows:
\begin{definition}\label{def:private_ldc_game}
A computationally bounded adversarial channel $\mathsf{C}$ with error rate $\rho$ is a probabilistic polynomial time algorithm which repeatedly interacts with the encoding algorithm $\Enc$ and the decoding algorithm
$\Dec$ polynomially many times until it terminates. Each iteration takes place as follows:
\begin{enumerate}
    \item Given a security parameter \secpar, the key generation algorithm outputs a secret key $\sk \leftarrow \mathsf{GenKey}(1^\secpar)$. The secret is given to both the sender (encoder) and the receiver (decoder) but not to the channel. The channel is given \secpar.
    \item In the $h\th$ iteration, the channel $\mathsf{C}$ chooses a message $x^{(h)} \in \{0, 1\}^k$ and hands it to the sender.
    \item The sender computes $y^{(h)} \leftarrow \Enc(x^{(h)}, \sk)$ and hands the codeword $y^{(h)} \in \{0,1\}^K$ back to the channel.
    \item The channel corrupts at most a fraction $\rho$ of all $K$ bits in $y^{(h)}$ to output the corrupted codeword $y'^{(h)}$, i.e., $\HAM(y^{(h)}, y'^{(h)}) \leq \rho K$. It gives $y'^{(h)}$ and a challenge bit $j$ to the receiver's \Dec
    \item The receiver makes at most $\ell$ (possibly adaptive) queries into the new codeword $y'^{(h)}$ and outputs $b \leftarrow \Dec(j, \sk)$.
\end{enumerate}
\end{definition}

Using such a setup, Ostrovsky~\etal give explict constructions of \emph{one-time} private \LDCs\ i.e. \LDCs\ that may be used for exactly one-round of communication in definition \ref{def:private_ldc_game}. 
They achieve this by \emph{code scrambling} (Appendix \ref{appendix:codescrambling}) a simple repetition code of the original message. 
Good locality is then achieved by reading and unscrambling only the bits exactly corresponding to indices of repetitions of the queried bit. 
It turns out that this simple code does not have good rate as the number of repetitions that need to be applied per symbol of the original message in the repetition code is not constant. 
Due to this, the authors employ a strategy where the message to be encoded is divided into blocks of small size. 
Each block is then encoded using an error correcting code of constant rate, and then the concatenation of all the encoded blocks is scrambled using the secret key. 
Good locality is achieved by reading and unscrambling only the bits exactly corresponding to the block containing the queried bit. 
Specifically, Ostrovsky~\etal give constructions of private \LDC $(\opskey,\opsenc,\opsdec)$ over the binary alphabet and show the following against adversarial channels:

\begin{theorem}[\cite{OPS}]\label{thm:ops-priv-loc-dec-code} Let $f(\secpar)$ be any function such that $f(\secpar) = \omega(\log\secpar)$. Then,
there exists a constant $\opsrho$ such that $(\opskey,\opsenc,\opsdec)$ is a \emph{one time} private $\opsquery$-locally decodable code with $\opsquery = f(\secpar)$ and constant information rate $(\beta_\mathsf{OPS})$ that correctly decodes from error rate $\opsrho$ with probability at least $1 - 2^{-\opsquery}$.
\end{theorem}

In Section \ref{sec:definitions}, we introduce Definition \ref{def:game-based-private-LDC} as an alternative to working with Definitions  \ref{def:private-LDCs} and \ref{def:private_ldc_game}. Furthermore, we make use of this alternative definition of private-\LDCs{} throughout the main sections. We thus present Theorem \ref{thm:ops-priv-loc-dec-code} in an alternative form where private \LDCs{} are presented as Definition \ref{def:game-based-private-LDC}. We use this in instantiating our framework in Section \ref{sec:framework} (Corollaries \ref{cor:p_constant_main} and \ref{cor:p_whp_main}).

\begin{theorem}[Alternative to Theorem \ref{thm:ops-priv-loc-dec-code}]
Let $f(\secpar)$ be any function such that $f(\secpar) = \omega(\log\secpar)$. Then, for security parameter \secpar and for all $K > k > 0$ such that $k = \poly(\secpar)$ where $\poly$ is any non-zero polynomial, there exists a $(K,k)_2$ coding scheme $\mathsf{C_{OPS}\parameters{K,k,\secpar}} = (\opskey,\opsenc,\opsdec)$ that is a one-time $(\opsquery, \opsrho, \probrecov_{\mathsf{OPS}}, \epsilon_{\mathsf{OPS}})-$private \LDC where $\opsquery = f(\secpar)$, $\opsrho$ is a constant, $\probrecov_{\mathsf{OPS}} = 1$, and $\epsilon_{\mathsf{OPS}} \leq \lenmsg \left(\frac{e}{4}\right)^{-\opsrho\opsquery}$ is negligible in the security parameter.
\end{theorem}



\end{document}